\documentclass[english,numberedappendix,letterpaper,iop,appendixfloats]{emulateapj}
\usepackage[T1]{fontenc}
\usepackage{babel}
\usepackage{amssymb}
\usepackage{graphicx}
\usepackage[unicode=true]
 {hyperref}
\usepackage{breakurl}

\makeatletter

\slugcomment{}
\shorttitle{}
\shortauthors{}

\makeatother

\begin{document}

\title{Multi-Zone Models of Superbursts from Accreting Neutron Stars}

\author{L.~Keek and A.~Heger}

\email{laurens@physics.umn.edu}

\affil{School of Physics and Astronomy, University of Minnesota, 116 Church
Street SE, Minneapolis, MN 55455, USA}
\begin{abstract}
Superbursts are rare and energetic thermonuclear carbon flashes observed
to occur on accreting neutron stars. We create the first multi-zone
models of \emph{series} of superbursts using a stellar evolution code.
We self-consistently build up the fuel layer at different rates, spanning
the entire range of observed mass accretion rates for superbursters.
For all models light curves are presented. They generally exhibit
a shock breakout, a precursor burst due to shock heating, and a two-component
power-law decay. Shock heating alone is sufficient for a bright precursor,
that follows the shock breakout on a short dynamical time scale due
to the fall-back of expanded layers. Models at the highest accretion
rates, however, lack a shock breakout, precursor, and the first power
law decay component. The ashes of the superburst that form the outer
crust are predominantly composed of iron, but a superburst leaves
a silicon-rich layer behind in which the next one ignites. Comparing
the model light curves to an observed superburst from 4U~1636-53,
we find for our accretion composition the best agreement with a model
at three times the observed accretion rate. We study the dependence
on crustal heating of observables such as the recurrence time and
the decay time scale. It remains difficult, however, to constrain
crustal heating, if there is no good match with the observed accretion
rate, as we see for 4U~1636-53.
\end{abstract}


\section{Introduction}

X-ray flares have been observed from accreting neutron stars that
are similar to Type I X-ray bursts, but that are a thousand times
more energetic and last up to a day. Normal bursts (e.g., \citealt{Lewin1993,Strohmayer2006})
result from hydrogen and helium burning to carbon and, through a series
of alpha captures, the \textsl{$\alpha$p}-process, and proton captures,
the \textsl{rp}-process, to heavier elements (\citealt{Schatz2001,2003Schatz}).
The long flares, named `superbursts', are attributed to the runaway
thermonuclear burning of carbon in a $100\,\mathrm{m}$ thick layer
of ashes of normal bursts (\citealt{Cumming2001,Strohmayer2002}).
The day long decay is explained as being the cooling time scale of
a layer of that thickness (\citealt{2004CummingMacBeth}). Because
it takes typically about one year to build up this $100$~meter thick
layer, superbursts are much rarer than regular bursts. The first superbursts
have been discovered relatively recently: it was only after the launch
of the BeppoSAX and RXTE observatories that enough exposure time was
collected to be able to detect such rare events (\citealt{Cornelisse2000,Strohmayer2002}).
At the time of writing, $20$ (candidate) superbursts have been observed
from 11 sources (see \citealt{Keek2008int..work} for an overview,
and \citealt{Kuulkers2009ATel}, \citealt{Chenevez2011ATel}, \citealt{Zand2011ATel}
for recent discoveries).

Two types of superbursts are discerned based on the composition of
the material that is accreted from the companion star. Most superbursters
are thought to accrete hydrogen-rich material. Their superbursts are
energetic, but the peak brightness does not reach the Eddington limit.
Superbursts have been observed from 4U~0614+91 (\citealt{Kuulkers2010})
and 4U~1820-30 (\citealt{Strohmayer2002}, and a candidate \citealt{Zand2011ATel}).
These sources are so-called ultra-compact X-ray binaries (UCXBs).
UCXBs have a binary period of less than $80$ minutes. In such a small
orbit, stable mass transfer by Roche-lobe overflow can only occur
from an evolved star that has lost its hydrogen envelope. The material
accreted onto the neutron star, therefore, contains no hydrogen, but
may contain helium. The superburst from 4U~$1820-30$ reached the
Eddington luminosity and displayed photospheric radius expansion.
For 4U~0614+91 the onset of the superburst was not observed.

Most superbursting sources have a high accretion rate $\dot{M}$ of
at least $10\%$ of the Eddington-limited rate $\dot{M}_{\mathrm{Edd}}$
at the time of the superburst. Exceptions are 4U~0614+91 with $\dot{M}\simeq0.01\dot{M}_{\mathrm{Edd}}$
(\citealt{Kuulkers2010}), and 4U~1608-522 where the accretion rate
at the time of the superburst was high, but where the average rate
over the previous years was $\dot{M}\simeq0.01\dot{M}_{\mathrm{Edd}}$
(\citealt{Keek2008}). The $\alpha$-parameter, the ratio of the accretion
fluence between normal bursts and the fluence of a burst, is typically
high: $\alpha\simeq1000$ (\citealt{Zand2003}). This indicates a
relatively large part of the accreted material is burned in a stable
manner instead of in bursts, and this may be necessary to achieve
high enough carbon fractions. No superbursts have been observed from
sources that only have stable burning and no bursts, though lower
limits on the possible recurrence time have been determined (\citealt{kee06}).
Although bursts reduce the carbon content of the envelope in the production
of heavy elements, it has been suggested that the heavy elements are
necessary for reducing the thermal conductivity, insuring that the
superburst ignition is reached at the observed depth in the envelope
(\citealt{Cumming2001}).

From fits of superburst-decay models to observed light curves (\citealt{2004CummingMacBeth}),
\citet{Cumming2006} deduce that superbursts ignite at a column depth
of $y\simeq10^{11}-10^{12}\,\mathrm{g\, cm^{-2}}$ in a layer with
a carbon mass fraction of $X_{12}\simeq20\%$. It is a challenge for
models to explain these ignition column depths. The carbon mass fractions
are higher than what one-dimensional models that include large nuclear
networks predict to be present in the ashes of normal bursts (\citealt{Woosley2004,Fisker2008}).
\citet{2005Cooper} suggest that the companion stars of superbursters
donate material with a CNO content that is four times higher than
solar.

Superbursts ignite close to the outer crust, and as such are sensitive
to the thermal properties of the crust, which are not yet well understood
(\citealt{2004Brown}). In turn, the temperature of the crust depends
on neutrino-cooling in the neutron star core, which is also ill-constrained.
Therefore, superbursts provide an observational measure of the thermal
properties of the outer crust, and constrain the physics in the crust
and the core (\citealt{Cumming2006,Page2005}).

The start of the superburst was observed only in eight cases. In six
of these, a short precursor burst is detected. For the other two superbursts,
the data was not of sufficient quality to exclude the presence of
a precursor, with the possible exception of 4U~1608-522, although
the detection of the superburst onset must be regarded tentative for
this source (\citealt{Keek2008}). \citet{Weinberg2007} explain the
precursor as the result of a shock generated by the superburst ignition.
This shock travels outwards through the envelope and triggers the
ignition of either a helium-rich layer or another carbon-rich layer.
The resulting flash is observable as the precursor burst.

In this paper we create a series of one-dimensional models of the
neutron star envelope, where for the first time we self-consistently
build up a carbon-rich layer at rates similar to the observed accretion
rates. We follow the carbon burning during several consecutive superbursts.
The dependence of observable properties of the bursts on crustal heating
is investigated. A possible hydrogen or helium-rich atmosphere is
not modeled in this paper.

\section{Neutron star envelope model}

\subsection{Stellar evolution code}

We employ the one-dimensional hydrodynamics stellar evolution code
KEPLER (\citealt{Weaver1978}). We use a version of KEPLER that differs
from the version used in recent studies (e.g., \citealt{Woosley2002RvMP,Woosley2004,Heger2007})
in the accretion scheme and the opacities that are used. We model
the neutron star envelope on a one-dimensional Lagrangian grid in
the radial direction, under the assumption of spherical symmetry.
The grid points represent the boundaries between concentric shells,
that each have a certain mass, chemical composition, temperature,
density, luminosity and radial velocity. Alternatively, a model could
be considered a local `wedge' of the neutron star, that would be well-approximated
in a plane-parallel geometry. 

Zones are added and removed in order to maintain an optimal grid for
resolving gradients in all quantities, such that temperature, density,
and radius vary from one zone to the next by at least 10\% and at
most 25\%. Furthermore, zones are not removed if they extend over
$0.02$ in $\log y$, where $y$ is the column depth. The effects
of different rezoning criteria was tested in a limited number of models;
the most important calculated properties such as burst recurrence
times and energetics varied by at most a few percent. The mass of
each zone as well as the size of each time step are recorded, such
that small values are not lost due to numerical precision.

We implicitly solve the equations of mass, energy, and momentum conservation
(\citealt{Weaver1978}). The equation of state allows for (non-)degenerate
and (non-)relativistic electrons.

To follow the chemical evolution we have the use of two networks of
nuclear reactions. An adaptive network follows a large number of reactions
among hundreds of isotopes (\citealt{Woosley2004}). Because this
network is computationally expensive, most of our calculations only
use an approximation network consisting of 19 isotopes (\citealt{Weaver1978}).
It includes the carbon fusion reactions as well as photodisintegration.
Comparison of superburst models created using either network shows
a $3.3\%$ shorter recurrence time for the model with the approximation
network, and a $3.1\%$ lower burst fluence. This indicates that the
large network generates $0.2\%$ more energy per unit mass than the
approximation network. There is no notable difference in the light
curves.

We take into account neutrino energy loss (\citealt{Itoh1996}), radiative
opacity (\citealt{Iben1975}), and electron conductivity (\citealt{Itoh2008}).

We consider convection using the Ledoux criterion, as well as semiconvection
and thermohaline mixing (e.g., \citealt{Heger2000}). The induced
mixing of the chemical composition is implemented as a diffusive process
using mixing-length theory (e.g., \citealt{Clayton1968book}). Rotation
and magnetic fields are not considered in these models.

\subsection{Accretion and decretion\label{sub:Accretion-and-decretion}}

Previous studies of X-ray bursts with the KEPLER code implemented
accretion by increasing the pressure at the outer zone over time to
simulate the build-up of a column of material (e.g., \citealt{Woosley1984,Taam1996}).
When this pressure reached a certain value, an extra zone containing
the accreted mass was added on top of the model. This induced a momentary
reduction of the time step as well as an artificial dip in the light
curve. In the present study we employ an improved accretion scheme
that solves these issues, allowing for larger time steps between subsequent
bursts, and producing light curves without the aforementioned artifacts.

Mass accretion is implemented by increasing the mass of one zone in
the model at each time step at the mass accretion rate. The zone is
selected at a pre-defined column depth such that it lies above the
region where thermonuclear burning takes place, but far enough below
the surface that the mass added to it is small compared to the layer
above (this avoids constant rezoning of the small surface zones).
Once the mass of the zone reaches a certain limit, it is split together
with one neighboring zone into three zones, conserving energy, momentum,
composition, and gradients. The chemical composition of the zone and
all zones above, up to the surface, is advected to account for the
composition of the accreted material. Furthermore, the radial positions
of the zones above the mass addition point are adjusted, and the energy
gained from compressional heating of the accreted material is taken
into account.

Increasing the mass of a model leads to increased neutrino emission
near the bottom. To avoid this we maintain a constant total mass for
the model, by decreasing the mass of the inner zone at the same rate
as at which mass is accreted. The radius of the inner boundary is
kept fixed, and all other zones are moved downward, such that the
density in the inner zone is conserved. Once the first zone's mass
is reduced below a certain limit, the three inner zones are merged
into two, again conserving energy, momentum, composition, and gradients.

\subsection{Substrate\label{sub:Substrate}}

The inner part of the models is formed by an iron substrate, on top
of which the carbon-rich superburst fuel is accreted. Heat generated
in a burst can diffuse into the substrate, and be released toward
the surface on a longer time scale. This ensures a correct long-term
light curve. We performed tests that show that the substrate should
contain at least an order of magnitude more mass than the burst ignition
column. At low accretion rates the ignition column depth is relatively
large, requiring the substrate to be located deeper. 

The substrate lies below the superburst ignition depth, and reaches
into the outer crust, where neutrino emission becomes increasingly
important. If we choose too thick a substrate, most of the luminosity
at the inner boundary will be dissipated as neutrinos. This is especially
a problem for models with high accretion rates, which have a relatively
high crustal heating and thence larger neutrino losses.

The wide range of ignition column depths and amounts of neutrino losses
pose constraints on the substrate mass that vary as a function of
the mass accretion rate. At the lowest rates we choose the substrate
to have a mass of $2\cdot10^{28}\,\mathrm{g}$, and $2\cdot10^{26}\,\mathrm{g}$
at the highest rates. As a test, we create several models with the
same accretion rate and varying substrate sizes. The changes in the
burst parameters such as the recurrence times is at most a few percent
for the selected substrate sizes.

\subsection{Crustal heating}

The amount of crustal heating of the envelope depends on the nuclear
reactions in the neutron star crust, the crust's thermal conductivity,
and on the neutrino cooling in that layer and in the core. The processes
in the crust are not calculated explicitely, but the resulting heating
of the envelope is emulated by a \emph{fixed} luminosity at the inner
boundary. For each model, we assume a certain heat flux per accreted
nucleon $Q_{\mathrm{b}}$. Combined with the accretion rate, it specifies
this luminosity.

The inner part of our models, the substrate, reaches into the crust,
and the luminosity that reaches the superbursting region is reduced
by neutrino emission. Because we wish our results to be independent
of our prescription of crustal neutrino cooling, we report an effective
$Q_{\mathrm{b}}$, that is corrected for neutrino emission in the
substrate.

\subsection{Relativistic corrections\label{sub:Relativistic-corrections}}

The code we employ uses Newtonian gravity (calculated for each zone),
whereas for neutron stars general relativistic (GR) corrections are
significant. To take these corrections into account, we can state
that our results are applicable to any combination of neutron star
mass and radius that give a GR gravitational acceleration equal to
the Newtonian acceleration employed by the code. The full details
of the GR corrections are available in Appendix~\ref{sec:General-relativistic-corrections}.
Here we give one example, but note that the results of the models
are valid for any combination of mass and radius that yields the same
value of the gravitational acceleration as used in this study. 

An input mass of $1.4\, M_{\odot}$ and radius of $10\,\mathrm{km}$
yield a local Newtonian gravitational acceleration throughout the
envelope of $g\simeq1.87\cdot10^{14}\,\mathrm{cm\, s^{-2}}$. Using
the same mass and a larger radius of $11.2\,\mathrm{km}$, one obtains
the same value of the gravitational acceleration, but now including
GR corrections. So the results of the Newtonian model are valid for
a GR model with increased radius. Because of the larger radius, the
luminosity from our model has to be increased as well, by a factor
$1.12^{2}\simeq1.26$. This mass and radius imply for an observer
at infinity a gravitational redshift of $1+z\simeq1.26$. The observed
luminosity is reduced by a factor $1.26$, and the observed ratio
of the accretion luminosity and the Eddington limit is scaled by a
factor $0.99$. The GR corrected global mass accretion rate for an
observer at inifinity is the same as the input (non-redshifted) accretion
rate $\dot{M}$.

The results presented in this paper do not contain these corrections
unless indicated otherwise (e.g., Sect.~\ref{sub:Comparison-to-4U1636-53}).

\subsection{Light curves\label{sub:Light-curves}}

Light curves are generated using the luminosity in the outer zone
(e.g., \citealt{Taam1996}). This zone typically extends orders of
magnitude in column depth deeper below the neutron star surface than
the photosphere. The surface zone, therefore, has a much longer thermal
time scale than the photosphere. For our models this typically means
that thermal diffusion cannot change the light curve faster than on
a thermal time scale of $\sim10^{-4}\,\mathrm{s}$ for an outer zone
of $10^{16}\,\mathrm{g}$. Dynamic processes such as shocks, however,
can heat the outer zone much faster, producing variations in the light
curve on shorter time scales. We do not correct for the time it would
take to transport heat through the outer zone to the `real' photosphere,
which is a reasonable approximation for the dynamic processes because
of the short spatial distance to the photosphere.

As explained in the previous subsection, no GR corrections are applied
to the light curves.

\subsection{Initial model setup\label{sub:Initial-model-setup}}

For the inner boundary we set the radius to $10\,\mathrm{km}$ and
the enclosed mass to $1.4\, M_{\odot}$ (using Newtonian gravity).
The outer boundary is initially set at a column depth of $y=10^{9}\,\mathrm{g\, cm^{-2}}$.
Once accretion is turned on, zones are quickly added such that the
boundary is at $y=10^{3}\,\mathrm{g\, cm^{-2}}$, which corresponds
to the outer zone having a mass of $\sim10^{16}\,\mathrm{g}$. Note
that we refrain from resolving the photosphere at $y\simeq1\,\mathrm{g\, cm^{-2}}$,
because this would require very light zones that display unphysical
behavior in the presence of shocks.

The model initially consists of the iron substrate (50 zones) and
five zones containing a mixture of $80\%$ $^{56}\mathrm{Fe}$ and
$20\%$ $\mathrm{^{12}C}$, which is later used as accretion composition.
This is the typical mass fraction of carbon \citet{Cumming2006} found
from fits to observed light curves of hydrogen accreting superbursters,
and $^{56}\mathrm{Fe}$ is the most abundant isotope in the ashes
of hydrogen-rich X-ray bursts (e.g., \citealt{Woosley2004}). Because
of the different heat sources (crustal, compressional heating) and
sinks (surface radiation, neutrino emission), the model must be brought
into thermal equilibrium before the simulation is started. The model
is evolved by the code over a period that is much longer than the
thermal time scale of any zone. In this period we do not consider
nuclear burning and mixing processes. With respect to accretion, we
do not change the mass and composition of the model, but we do advect
compressional heating throughout all zones. Crustal heating is applied
as well. Once the model is in thermal equilibrium, we reset the simulation
time to zero, and enable accretion fully, as well as nuclear burning
and mixing processes. Because of the accretion of mass, zones are
added: a typical model contains approximately 400 zones, about 50
of which are located in the substrate, and another 50 form the outer
region where accretion is implemented.

We create models with different values for $Q_{\mathrm{b}}$ and for
the mass accretion rate $\dot{M}$, expressed as a fraction of the
Eddington limited rate $\dot{M}_{\mathrm{Edd}}$. In this paper we
use the Eddington limit for an atmosphere of solar composition on
a neutron star of $1.4\, M_{\odot}$ with a $11.2\,\mathrm{km}$ radius,
which corresponds to an Eddington luminosity of $L_{\mathrm{Edd}}=2.5\cdot10^{38}\,\mathrm{erg\, s^{-1}}$
and accretion rate of $\dot{M}_{\mathrm{Edd}}=1.96\cdot10^{-8}\, M_{\odot}\,\mathrm{year^{-1}}$.
Taking into account the gravitational redshift, the observed values
at infinity are $L_{\mathrm{Edd}\,\infty}=1.6\cdot10^{38}\,\mathrm{erg\, s^{-1}}$
and $\dot{M}_{\mathrm{Edd}\,\infty}\simeq1.56\cdot10^{-8}\, M_{\odot}\,\mathrm{year^{-1}}$.

\section{Results}

\subsection{Stable/unstable ignition}

\begin{figure}
\includegraphics{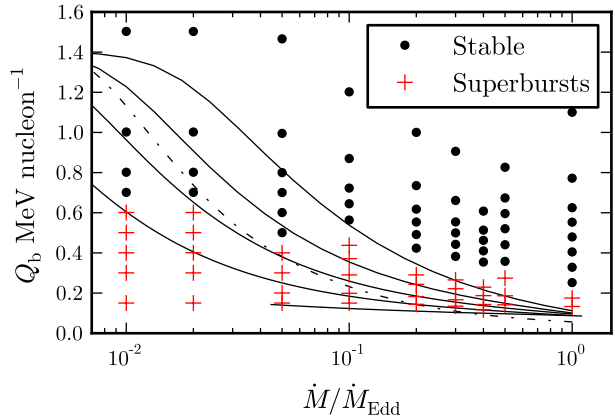}

\caption{\label{fig:Mass-accretion-rate}Mass accretion rate $\dot{M}$ in
units of the Eddington-limited rate, $\dot{M}_{\mathrm{Edd}}$, and
effective crustal heating parameter, $Q_{\mathrm{b}}$, for 78 models.
For each model we indicate whether carbon burning proceeds in a stable
manner or as a superburst. The lines correspond to different models
for crustal heating (Fig.~18 from \citealt{Cumming2006}): models
with higher neutrino cooling have lower values of $Q_{\mathrm{b}}$
(\emph{solid lines}); a model with a highly impure crust (\emph{dot-dashed
line}).}

\end{figure}

We create a series of models for different values of the mass accretion
rate and crustal heating. We vary the mass accretion rate in the range
where superbursts are observed, $0.01\leq\dot{M}/\dot{M}_{\mathrm{Edd}}\leq1.00$
(e.g., \citealt{Keek2008int..work}), and we vary the amount of crustal
heating between the minimum and maximum values suggested in the literature:
$0.1\leq Q_{\mathrm{b}}\leq1.5$ (e.g., \citealt{Haensel2003,Cumming2006,Gupta2007}).
In Fig.~\ref{fig:Mass-accretion-rate} we indicate for which values
of these parameters the models exhibit superbursts or stable carbon
burning. We include the different predictions for $Q_{\mathrm{b}}$
as a function of $\dot{M}$ from \citet{Cumming2006}, with the exception
of the model that includes Cooper pairs, as its neutrino emissivity
was shown to be overestimated (\citealt{Leinson2006}).

\begin{figure}
\includegraphics{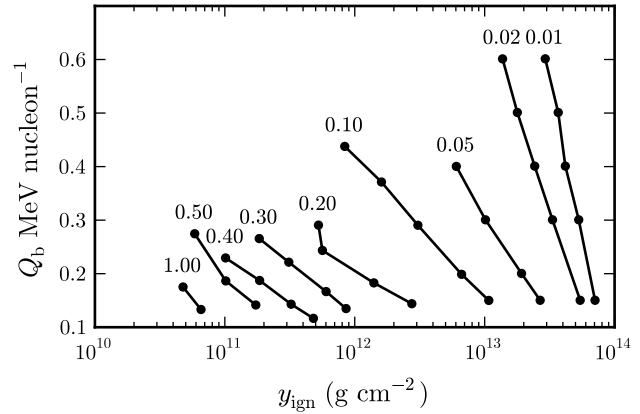}

\caption{\label{fig:yign_qb}Superburst ignition column depth $y_{\mathrm{ign}}$
as a function of the crustal heating parameter $Q_{\mathrm{b}}$ for
series of models with a fixed mass accretion rate indicated as a fraction
of $\dot{M}_{\mathrm{Edd}}$.}
\end{figure}
 
\begin{figure}
\includegraphics{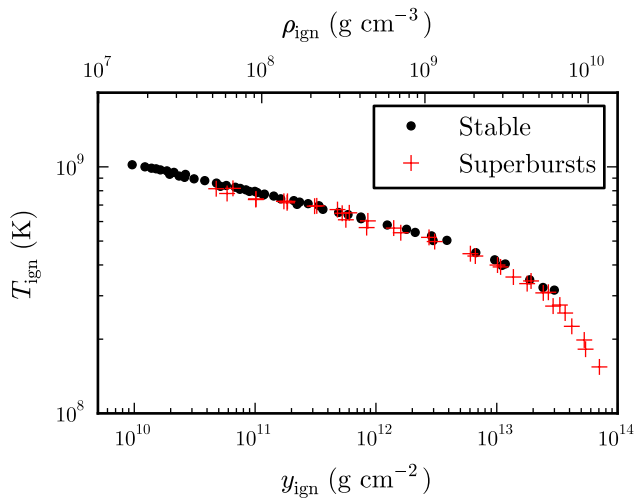}

\caption{\label{fig:yign_tign}Ignition (burning) temperature $T_{\mathrm{ign}}$,
density $\rho_{\mathrm{ign}}$ and column depth $y_{\mathrm{ign}}$
for models of superbursts (stable carbon burning), in a wide range
of mass accretion rates and crustal heating. The relation between
$\rho_{\mathrm{ign}}$ and $y_{\mathrm{ign}}$ used here (a power-law
fit to the numerical results) is accurate up to $1.2\%$.}
\end{figure}

Ignition occurs in our models at a column depth of $y_{\mathrm{ign}}\simeq10^{10}\,\mathrm{g\, cm^{-2}}$
for the models with the highest accretion rate and crustal heating,
and at $y_{\mathrm{ign}}\simeq7\cdot10^{13}\,\mathrm{g\, cm^{-2}}$
for the coolest models with the lowest accretion rate (Fig.~\ref{fig:yign_qb},
\ref{fig:yign_tign}). We determine $y_{\mathrm{ign}}$ in our models
from the location of the peak temperature just before the start of
the runaway (in case of unstable burning), or the peak energy generation
rate (in case of steady-state burning), which in both cases identifies
the bottom of the carbon-rich layer.

\begin{figure}
\includegraphics{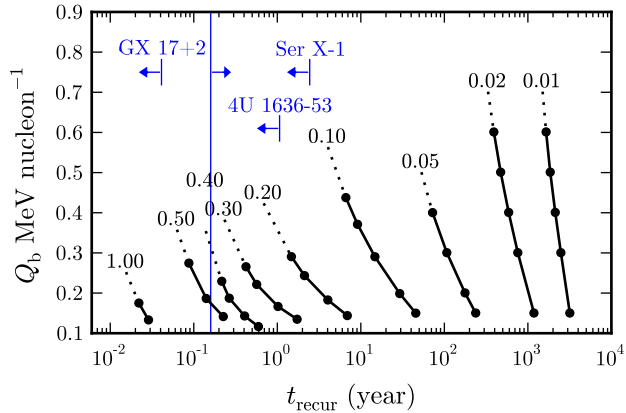}

\caption{\label{fig:trecur_qb}Superburst recurrence time, $t_{\mathrm{recur}}$,
as a function of the crustal heating parameter, $Q_{\mathrm{b}}$,
for series of models with a fixed mass accretion rate, indicated as
a fraction of $\dot{M}_{\mathrm{Edd}}$. The dotted extrapolations
continue after the hottest bursting models to the $Q_{\mathrm{b}}$
value where the first stable model was found in our grid. For three
sources we indicate the shortest observed recurrence time (at arbitrary
$Q_{\mathrm{b}}$). The average lower limit from BeppoSAX WFC data
for nine sources is indicated by a vertical line (\citealt{kee06}).}
\end{figure}

The bursting models exhibit recurrence times of several days up to
thousands of years (Fig.~\ref{fig:trecur_qb}). A given recurrence
time can be reproduced by a relatively hot model with a certain accretion
rate or a colder model with a somewhat higher accretion rate.

\subsection{Thermonuclear burning}

\begin{figure}
\includegraphics{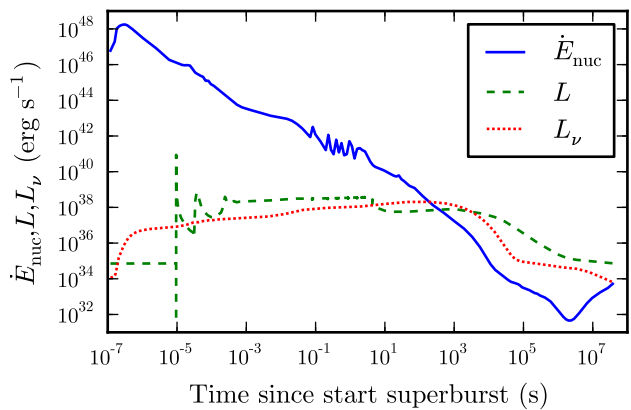}

\caption{\label{fig:energy}Energy generation rate, $\dot{E}_{\mathrm{nuc}}$,
surface luminosity, $L$, and neutrino luminosity, $L_{\nu}$, as
a function of time from the start of one superburst up to the onset
of the next, for a model with $\dot{M}=0.30\,\dot{M}_{\mathrm{Edd}}$
and $Q_{\mathrm{b}}=0.13\,\mathrm{MeV\, nucleon^{-1}}$.}
\end{figure}

To illustrate the thermonuclear burning processes during a superburst
we consider a model with $\dot{M}=0.30\,\dot{M}_{\mathrm{Edd}}$ and
$Q_{\mathrm{b}}=0.13\,\mathrm{MeV\, nucleon^{-1}}$. The energy generation
rate is highest $\sim2\cdot10^{-7}\,\mathrm{s}$ after the thermonuclear
runaway, and decreases roughly as $t^{-1}$ over the course of $\sim10^{6}\,\mathrm{s}$
(Fig.~\ref{fig:energy}). Afterwards, the energy generation rate
rises again due to increased carbon burning in a newly accreted fuel
layer, leading up to the next superburst.

\begin{figure}
\includegraphics{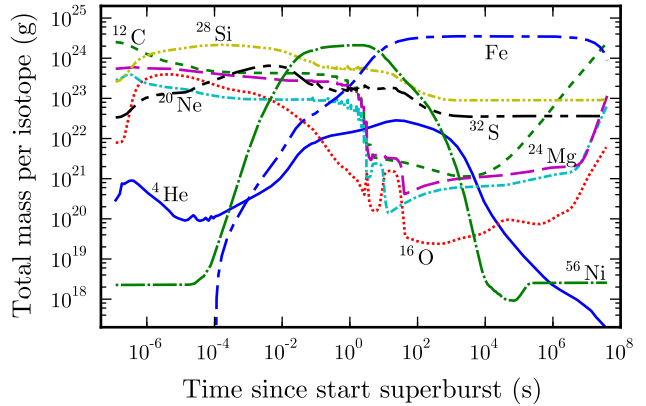}

\caption{\label{fig:masses}Total mass for a selection of isotopes as a function
of time from the start of one superburst up to the onset of the next,
for a model with $\dot{M}=0.30\,\dot{M}_{\mathrm{Edd}}$ and $Q_{\mathrm{b}}=0.13\,\mathrm{MeV\, nucleon^{-1}}$.
For Fe the mass at $t=0$ is subtracted: only the produced mass is
shown. The Fe mass decreases toward the end because of mass removal
at the inner zone of the model (Sect.~\ref{sub:Accretion-and-decretion}),
\emph{not} because of nuclear reactions.}
\end{figure}
 At the superburst onset a large fraction of the available carbon
burns through the $\mathrm{^{12}C(^{12}C,\alpha)^{20}Ne}$ reaction
(Fig.~\ref{fig:masses}). Subsequent $\alpha$-capture reactions
produce heavier isotopes, such as $\mathrm{^{24}Mg}$, $\mathrm{^{28}Si}$,
and $\mathrm{^{32}S}$. Photodisintegration causes the release of
more $\alpha$-particles, whose captures create iron-group elements
(e.g., iron and nickel). Electron-captures onto nickel produce iron,
which is the most abundant element in the superburst ashes after several
seconds. Note that in the employed approximation network the only
iron isotope is $\mathrm{^{54}Fe}$, whereas calculations with a large
network confirm that $\mathrm{^{56}Fe}$ is the most abundant isotope.

\begin{figure}
\includegraphics{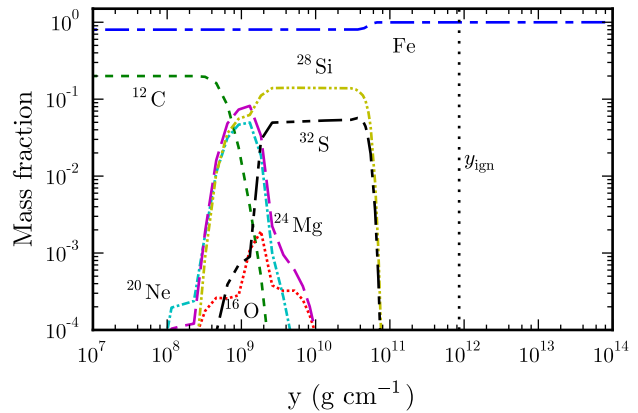}

\caption{\label{fig:massfractions}Mass fraction per isotope as a function
of column depth, $y$, approximately $10^{4}\,\mathrm{s}$ after the
start of a superburst for a model with $\dot{M}=0.30\,\dot{M}_{\mathrm{Edd}}$
and $Q_{\mathrm{b}}=0.13\,\mathrm{MeV\, nucleon^{-1}}$. The region
below $y=10^{7}\,\mathrm{g\, cm^{-2}}$ is omitted, as it has constant
carbon and iron fractions. The column depth at which the superburst
ignited, $y_{\mathrm{ign}}$, is indicated by the vertical dotted
line.}
\end{figure}
 Approximately $10^{4}\,\mathrm{s}$ after the burst start, the total
amount of carbon increases again when accretion adds carbon faster
than residual burning can take it away (Fig.~\ref{fig:masses}).
At this time there is a layer of pure iron directly above the ignition
depth, that accounts for over $90\%$ of the mass of the superbursting
layer (Fig.~\ref{fig:massfractions}). This is the composition that
forms the outer crust. In the outer part of the envelope, photodisintegration
was less efficient due to the lower temperature, resulting in ashes
that are more rich in $\mathrm{^{20}Ne}$, $\mathrm{^{24}Mg}$, $\mathrm{^{28}Si}$,
and $\mathrm{^{32}S}$. The new fuel piles on top of that layer. During
the next superburst these isotopes burn to iron group elements. Note
that models with stable carbon burning do not undergo photodisintegration,
and there the outer crust will be enriched with $\mathrm{^{20}Ne}$,
$\mathrm{^{24}Mg}$, $\mathrm{^{28}Si}$, and $\mathrm{^{32}S}$.

\begin{figure}
\includegraphics{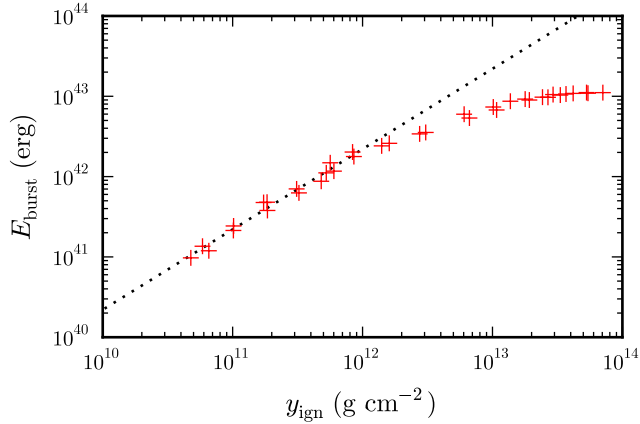}

\caption{\label{fig:yign_eb}Ignition column depth, $y_{\mathrm{ign}}$, and
the total burst energy, $E_{\mathrm{burst}}$, emitted at the surface
for all models. The linear relation below $y_{\mathrm{ign}}\lesssim10^{12}\,\mathrm{g\, cm^{-2}}$
is indicated by the dotted line.}
\end{figure}
Part of the generated energy leaves the envelope as photons from the
surface, and part is lost in neutrinos (Fig.~\ref{fig:energy}).
Neutrino emission is strongest at larger column depths in the substrate.
The total burst energy emitted in photons at the surface, $E_{\mathrm{burst}}$,
for all bursting models follows a linear relation for ignition below
$y_{\mathrm{ign}}\lesssim10^{12}\,\mathrm{g\, cm^{-2}}$ (Fig.~\ref{fig:yign_eb}).
At larger depths $E_{\mathrm{burst}}$ drops below this relation,
because an increasing part of the energy is emitted as neutrinos from
the substrate (the crust). The maximum $E_{\mathrm{burst}}$ in these
models is $1.1\cdot10^{43}\,\mathrm{erg}$.

\subsection{Shock and mixing}

\begin{figure}
\includegraphics[angle=90,width=246.0pt]{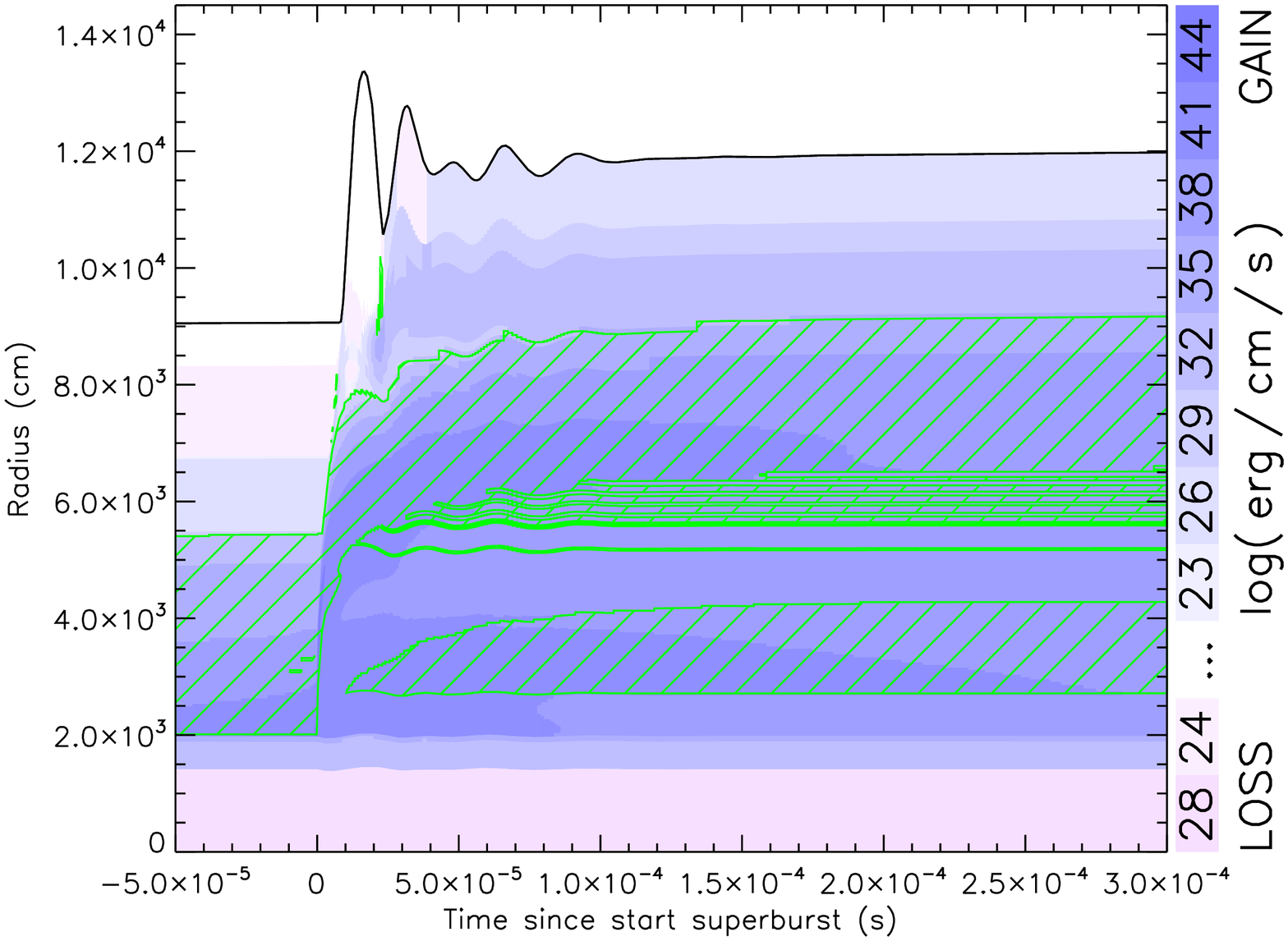}

\includegraphics[angle=90,width=246.0pt]{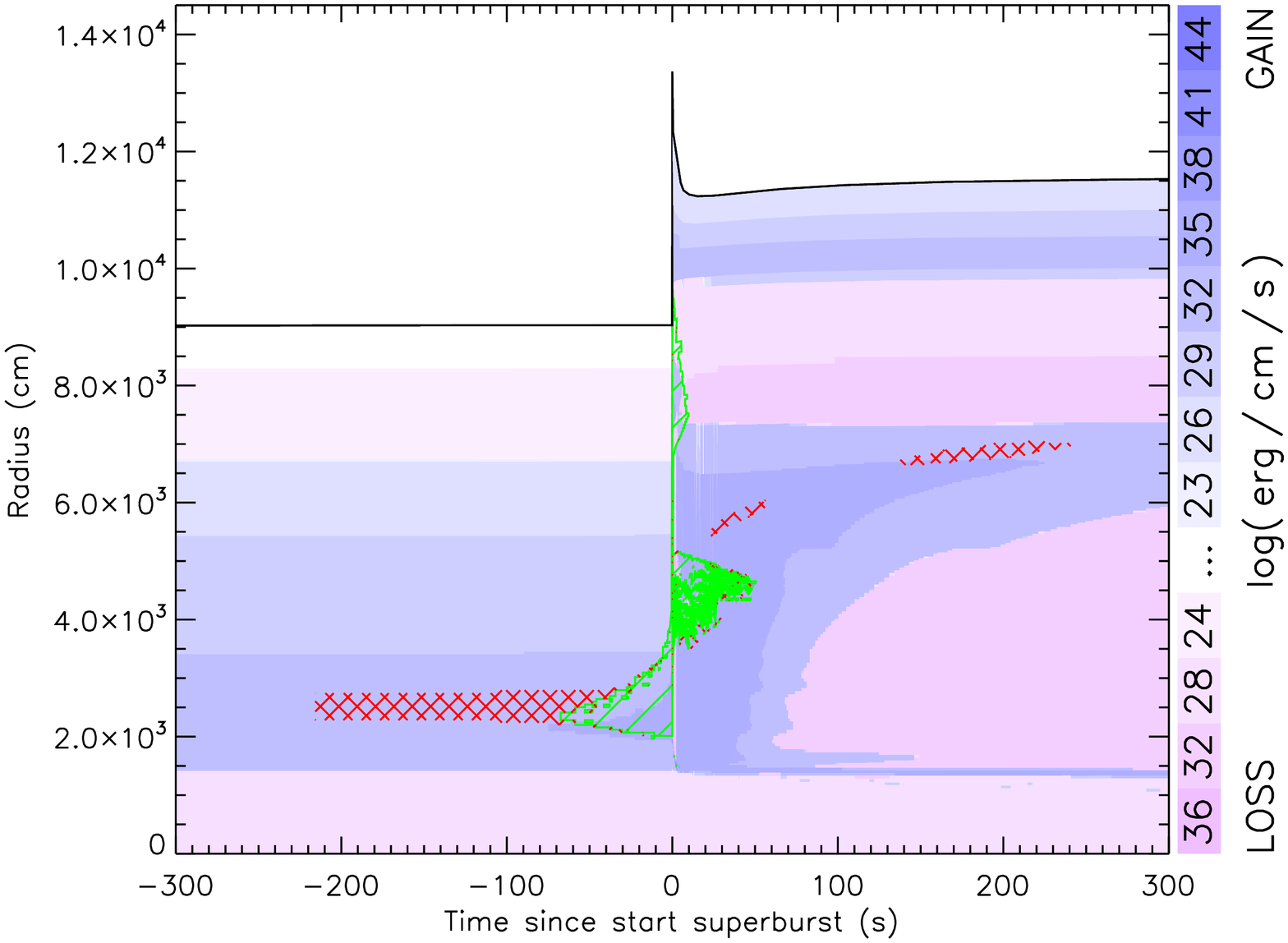}

\includegraphics[angle=90,width=246.0pt]{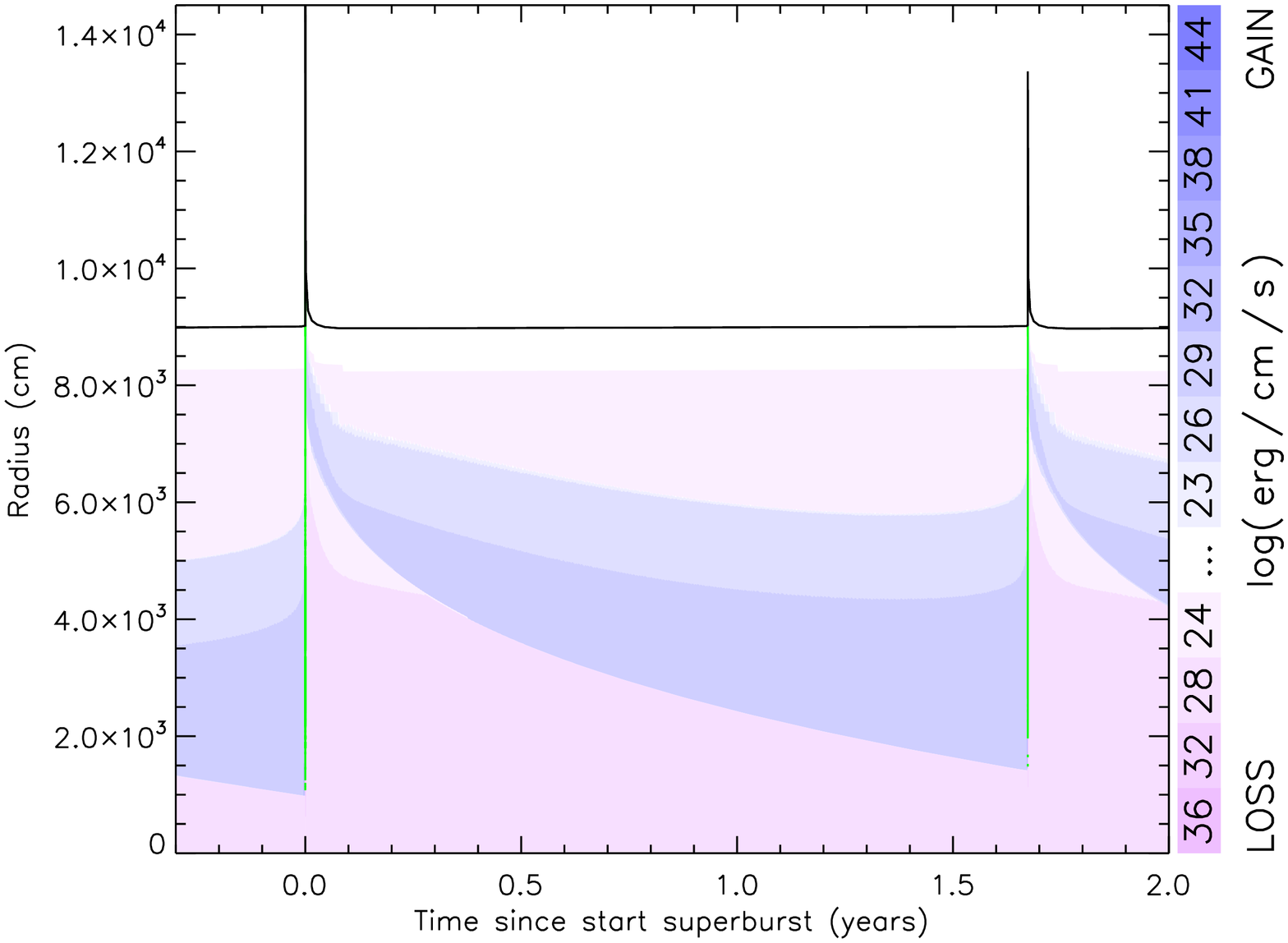}

\caption{\label{fig:convplot_0.3Edd}On three time scales: energy generation/loss
(color scale) in the neutron star envelope as a function of time in
a short interval around the superburst onset for a model with $\dot{M}=0.30\,\dot{M}_{\mathrm{Edd}}$
and $Q_{\mathrm{b}}=0.13\,\mathrm{MeV\, nucleon^{-1}}$. Green hatching
indicates convection, red cross hatching semiconvection.}
\end{figure}
 To study the hydrodynamic processes during a superburst we again
consider the model with $\dot{M}=0.30\,\dot{M}_{\mathrm{Edd}}$ and
$Q_{\mathrm{b}}=0.13\,\mathrm{MeV\, nucleon^{-1}}$. When the thermonuclear
runaway occurs at the bottom of the carbon-rich layer, the burning
initially proceeds as a detonation. A combustion wave moves outward,
creating a shock. After several microseconds the combustion wave slows
down, and burning spreads as a deflagration. The shock continues to
travel toward the surface on a microsecond time scale (Fig.~\ref{fig:convplot_0.3Edd}
top). The top layers are pushed outward, and subsequently fall back
on a dynamical time scale of approximately $10^{-5}\,\mathrm{s}$.
Afterwards the surface undergoes a damped oscillation. In Fig.~\ref{fig:convplot_0.3Edd}
we only show the envelope down to $y\simeq10^{8}\,\mathrm{g\, cm^{-2}}$,
which provides the most insight into the dynamic processes (see also
Sect.~\ref{sub:Precursor-burst}).

As the outer layers fall back, most of the kinetic energy is dissipated
into heat at a depth of $y\simeq10^{8}$ to $10^{10}\,\mathrm{g\, cm^{-2}}$.
Heating by the shock and the fall-back induces some carbon burning
in this region, leading to two regions of thermonuclear burning (Fig.~\ref{fig:convplot_0.3Edd}
top).

Several 100 seconds before and after the superburst onset, convection
mixes the composition in the envelope (Fig.~\ref{fig:convplot_0.3Edd}
middle). Briefly, at the thermonuclear runaway, the convective region
reaches close to the surface. After the burst, burning continues at
a very low rate at the bottom of the freshly accreted layer (Fig.~\ref{fig:convplot_0.3Edd}
bottom). During this time no convective mixing takes place. Once the
ignition column depth is reached, the next superburst occurs.

The compositional gradient created by the superburst induces thermohaline
mixing in the envelope. This mixing is, however, many orders of magnitude
slower than that due to convection at the time of burst onset.

\subsection{Light curve\label{sub:Light-curve}}

\begin{figure*}
\begin{centering}
\includegraphics{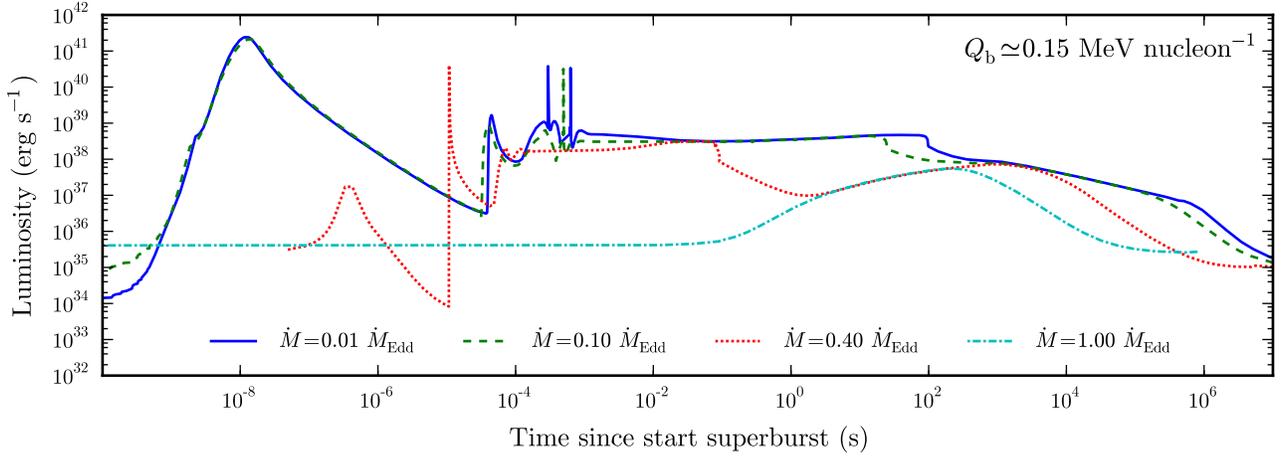}
\par\end{centering}

\caption{\label{fig:profile_qb}Light curves of superburst models with $Q_{\mathrm{b}}\simeq0.15\,\mathrm{MeV\, nucleon^{-1}}$
for a wide range of mass accretion rates.}
\end{figure*}
We generate light curves for all bursting models (Fig.~\ref{fig:all-light-curves}).
We compare a selection of light curves by taking the coldest model
($Q_{\mathrm{b}}\simeq0.15\,\mathrm{MeV\, nucleon^{-1}}$) in a wide
range mass accretion rates (Fig.~\ref{fig:profile_qb}). The curves
consist of several components: a shock breakout peak, a precursor,
a transition to the superburst peak, followed by a two-part power
law decay. Not every model exhibits each component. After the superburst
peak, the decay proceeds as $t^{-0.21}$. Following the break, the
decay steepens to $t^{-1.36}$. The models with the lowest mass accretion
rates, which are the coldest models with the largest ignition column
depth, have the longest decays. The $t^{-0.21}$ power law forms an
upper bound to the light curve. Toward higher accretion rates, the
time spent in the $t^{-0.21}$ part is smaller, until it becomes absent
at the highest rates, and there is a direct transition from the peak
to the $t^{-1.36}$ decay. 

Whereas the models with lower accretion rate exhibit a precursor,
at the highest accretion rates --- the hottest models --- it is absent.
Colder models have longer precursor bursts of up to $\sim10^{2}\,\mathrm{s}$.
We find precursors as short as $\sim10^{-1}\,\mathrm{s}$. All precursors
reach the Eddington limit and cause radius expansion. Depending on
the duration of the precursor, the transition to the superburst `peak'
around $\sim10^{3}\,\mathrm{s}$ can exhibit a drop in luminosity
below the peak value. The models with longer precursors lack this
dip.

\begin{figure}
\includegraphics{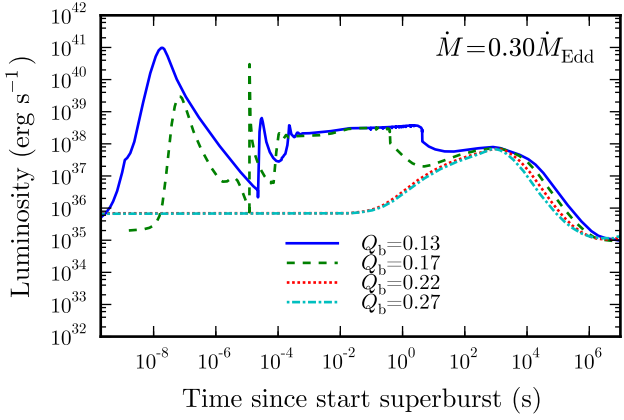}

\caption{\label{fig:profile_mdot_0.3}Light curves of superburst models with
$\dot{M}=0.30\,\dot{M}_{\mathrm{Edd}}$ for indicated values of $Q_{\mathrm{b}}$
in units of $\mathrm{MeV\, nucleon^{-1}}$.}
\end{figure}
To study the effect of crustal heating on the light curve, we compare
a series of light curves of simulations with the same mass accretion
rate, $\dot{M}=0.30\,\dot{M}_{\mathrm{Edd}}$, but with increasing
$Q_{\mathrm{b}}$ (Fig.~\ref{fig:profile_mdot_0.3}). For increasing
$Q_{\mathrm{b}}$, the duration of the superbursts increases from
$2.0\cdot10^{5}\,\mathrm{s}$ to $6.7\cdot10^{5}\,\mathrm{s}$, and
the time spent in the $t^{-0.21}$ phase reduces. The hottest models
again lack the precursor. Comparing the models with $Q_{\mathrm{b}}=0.13$
and $Q_{\mathrm{b}}=0.17$, the latter has a shorter duration precursor,
as well as a deeper dip in the light curve between the precursor and
the main peak after 1000~s.

Some light curves show a shorter precursor phase than expected, exhibiting
instead a small bump immediately after the precursor (e.g., some models
with $\dot{M}=0.10\,\dot{M}_{\mathrm{Edd}}$ in Fig.~\ref{fig:all-light-curves}).
This is due to the relaxation of the outer atmosphere following the
end of the radius expansion phase, and may be attributed to poor resolution
at the surface of the models. It is not caused by burning or mixing
processes.

\begin{figure}
\includegraphics{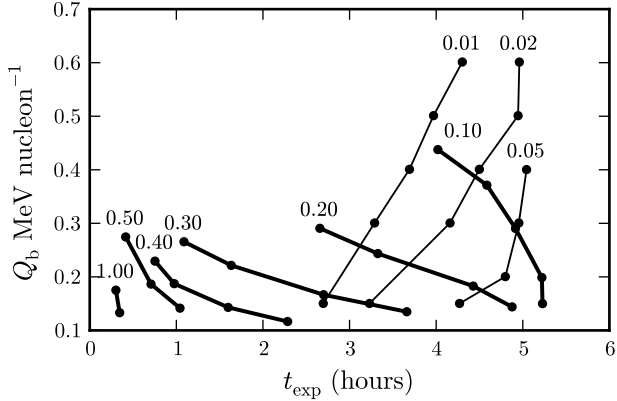}

\caption{\label{fig:tefold_qb}Exponential decay time, $t_{\mathrm{exp}}$,
vs. crustal heating, $Q_{\mathrm{b}}$, for series of models with
a certain mass accretion rate, indicated as a fraction of $\dot{M}_{\mathrm{Edd}}$.}
\end{figure}
We define the exponential decay time, $t_{\mathrm{exp}}$, of our
light curves as the time it takes from the superburst peak around
$t=10^{3}\,\mathrm{s}$ to drop one e-fold in luminosity (Fig.~\ref{fig:tefold_qb}).
It ranges from 18~minutes to 5.2~hours.

\subsection{Precursor burst\label{sub:Precursor-burst}}

\begin{figure}
\includegraphics{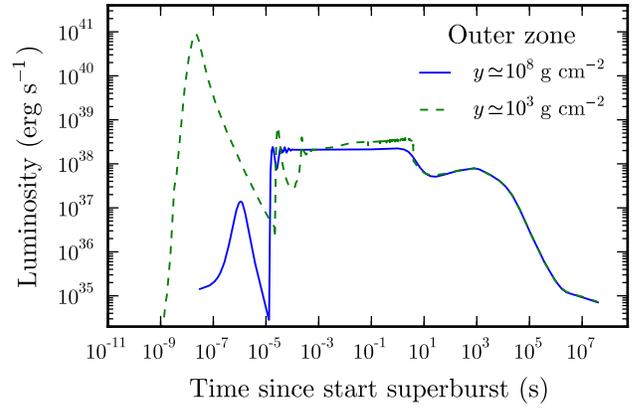}

\caption{\label{fig:atmospheres}Light curves of superburst models with $\dot{M}=0.30\,\dot{M}_{\mathrm{Edd}}$
and $Q_{\mathrm{b}}=0.13\,\mathrm{MeV\, nucleon^{-1}}$ with different
column depths for the outer zone.}
\end{figure}
As noted by \citet{Weinberg2007}, the overpressure of the shock is
larger at lower column depths. Hence, at lower column depths the shock
induces more heating and a larger radius expansion. To illustrate
this, we compare the light curves of two models that have both $\dot{M}=0.30\,\dot{M}_{\mathrm{Edd}}$
and $Q_{\mathrm{b}}=0.13\,\mathrm{MeV\, nucleon^{-1}}$, but one model
extends to a column depth of $y\simeq10^{8}\,\mathrm{g\, cm^{-2}}$,
and the other to $y\simeq10^{3}\,\mathrm{g\, cm^{-2}}$ (Fig.~\ref{fig:atmospheres}).
The model with the more extended envelope has a shock breakout peak
that has an approximately 50 times faster rise and a super-Eddington
luminosity of $0.89\cdot10^{41}\,\mathrm{erg\, s^{-1}}$, whereas
the other model's shock peak reaches only $1.4\cdot10^{37}\,\mathrm{erg\, s^{-1}}$.
After the shock breakout, the model with the more extended envelope
has stronger radius expansion and larger variations in luminosity.
The latter reaches a $1.7$ times higher value than for the other
model. The stronger shock heating leads to a lower opacity, which
increases the Eddington limit, allowing for higher surface luminosities.
4 seconds after the start of the superburst, the radius expansion
phase ends. The more extended atmosphere model displays a sharper
drop in luminosity. The smoother luminosity decrease of the other
model may lead to the interpretation that the precursor in this case
has a duration that is several seconds shorter. After the precursor,
the light curves of the two models are virtually identical. Therefore,
both the duration and the peak luminosity of the shock breakout and
the precursor depend greatly on the extend of the atmosphere. Note
that the neutron star photosphere is expected at a column depth of
approximately $y\simeq1\,\mathrm{g\, cm^{-2}}$, and it is likely
that at that column depth the shock breakout peak as well as the precursor
properties are different from our results.

The shock heating of the outer layers induces some carbon burning,
but the amount of heat generated by the nuclear reactions is too small
to substantially alter the light curve, as we checked by comparing
to a model where burning was disabled in that region. Therefore, the
precursor is in these models virtually completely powered by shock
heating.

\begin{figure}
\includegraphics[angle=90,width=1\columnwidth]{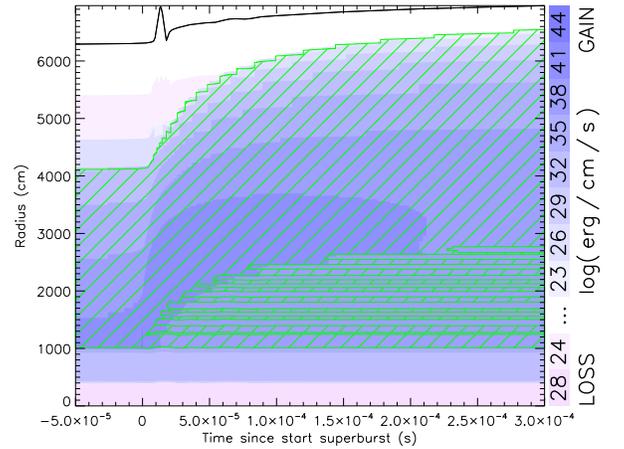}

\caption{\label{fig:convplot_0.3Edd-1}For a model without a precursor ($\dot{M}=0.30\,\dot{M}_{\mathrm{Edd}}$
and $Q_{\mathrm{b}}=0.22\,\mathrm{MeV\, nucleon^{-1}}$) the energy
generation/loss (color scale) in the neutron star envelope as a function
of time in a short interval around the superburst onset. Green hatching
indicates convection.}
\end{figure}

The hottest models at the highest accretion rates do not show precursors
at all (Sect.~\ref{sub:Light-curve}). We compare the model with
a precursor from Fig.~\ref{fig:convplot_0.3Edd} to a hotter one
($Q_{\mathrm{b}}=0.22\,\mathrm{MeV\, nucleon^{-1}}$) that lacks a
precursor (Fig.~\ref{fig:convplot_0.3Edd-1}). The hotter model's
burst has a shallower ignition depth, and is, consequently, less powerful.
The shock causes only minimal radius expansion, and does not provide
enough heating to produce a precursor burst or to ignite carbon close
to the surface. In this case there is only one region of carbon burning
(Fig.~\ref{fig:convplot_0.3Edd-1}).

\subsection{Comparison to 4U~1636-53\label{sub:Comparison-to-4U1636-53}}

\begin{figure}
\includegraphics{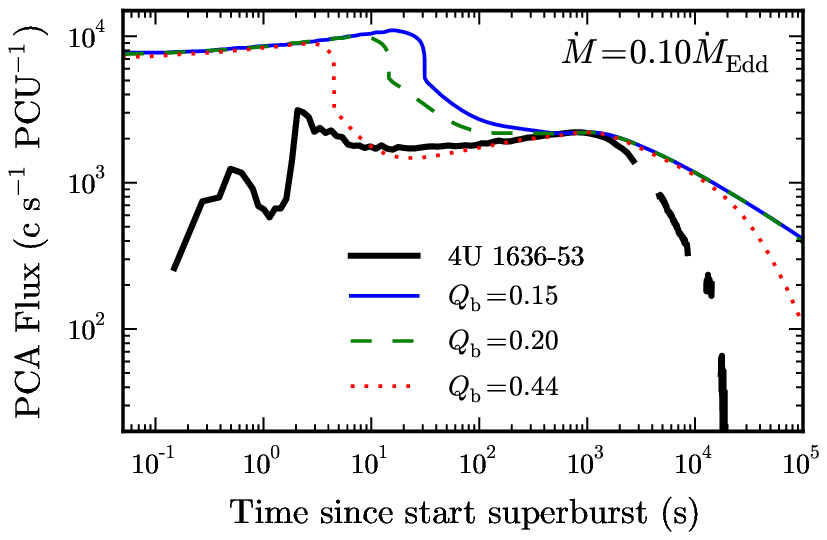}

\includegraphics{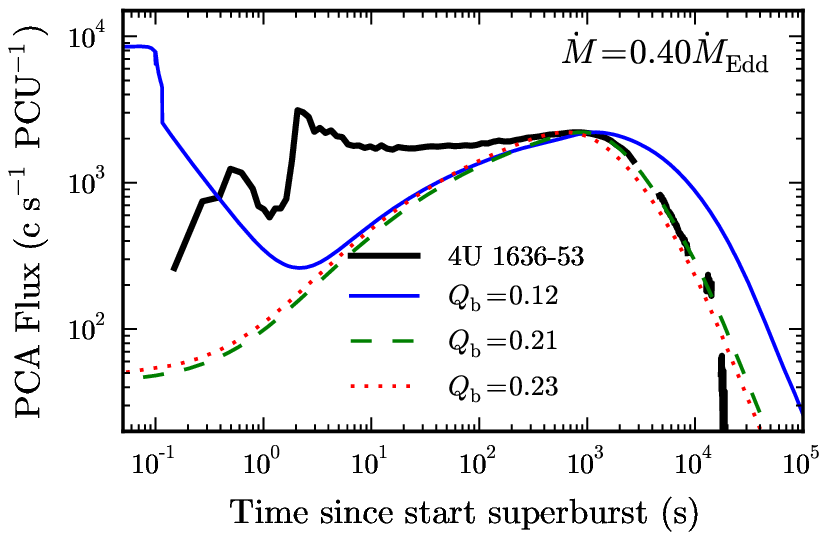}

\includegraphics{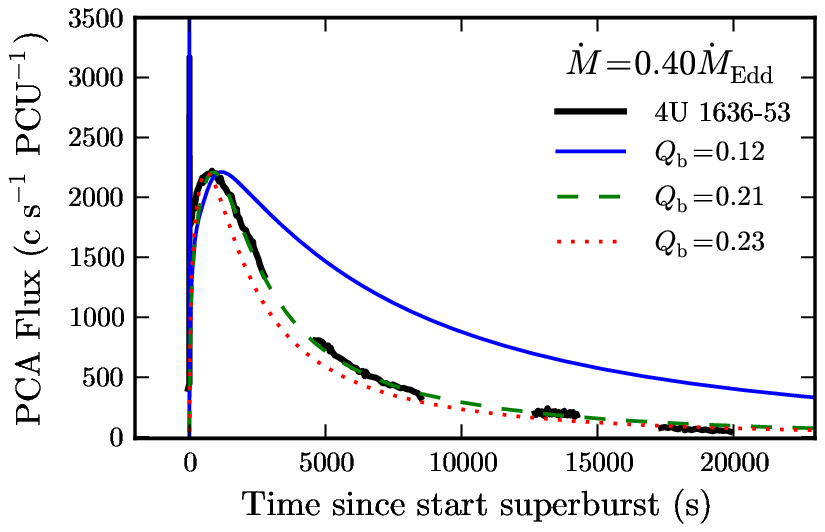}

\caption{\label{fig:profile_1636}Comparison of the observed light curve of
4U~1636-53 to light curves of superburst models with $\dot{M}=0.10\,\dot{M}_{\mathrm{Edd}}$
and $\dot{M}=0.40\,\dot{M}_{\mathrm{Edd}}$ for indicated values of
$Q_{\mathrm{b}}$ on a logarithmic scale. The bottom figure shows
the data of the middle one on a linear scale.}
\end{figure}

We compare several of our simulated light curves to the observed light
curve from the superburst of \object{4U~1636-53}, that was observed
with the Proportional Counter Array (PCA) on-board the Rossi X-ray
Timing Explorer (RXTE) (Fig.~\ref{fig:profile_1636}; \citealt{Strohmayer2002a};
see also \citealt{Kuulkers2003a,2004Kuulkers}). The PCA consists
of five Proportional Counter Units (PCUs). We use standard 1 data
from PCU 2 in the full instrument band-pass, and correct it for dead
time following the prescription from the RXTE Cookbook%
\footnote{\href{http://heasarc.gsfc.nasa.gov/docs/xte/recipes/pca_deadtime.html}{http://heasarc.gsfc.nasa.gov/docs/xte/recipes/pca\_{}deadtime.html}%
}. We subtract the persistent emission as measured from the end of
the last orbit that we consider. In this we assume the persistent
flux to remain constant during the superburst, which is probably not
the case: in the day preceding the superburst, the persistent flux
varied by around $10^{2}\,\mathrm{counts\, s^{-1}\, PCU^{-1}}$. Because
the superburst lasts longer than the orbit of RXTE, the observation
was interrupted three times by Earth occultations, resulting in gaps
in the light curve.

The simulated light curves are constructed taking into account the
PCA's instrument response and astrophysical effects. We use the surface
radius and temperature from our models to calculate the blackbody
emission from the superbursts. The temperature is increased by a typical
color correction factor of $1.5$ to account for deviation from a
pure blackbody spectrum due to Compton scattering close to the neutron
star surface (e.g., \citealt{Suleimanov2010}). We apply a gravitational
redshift of $1+z=1.26$ for a neutron star with a gravitational mass
of $1.4\, M_{\odot}$ and a radius of $11.2\,\mathrm{km}$ in the
local rest frame (see Sect.~\ref{sub:Relativistic-corrections};
Appendix~\ref{sec:General-relativistic-corrections}). The effect
of interstellar absorption by hydrogen is taken into account using
the model by \citet{1983Morrison}, that uses solar abundances from
\citet{Anders1982}, using a hydrogen column of $2.5\cdot10^{21}\,\mathrm{cm^{-2}}$
(\citealt{Asai2000}). We take into account the effective area of
the PCUs at different energies using the table provided with the software
package PIMMS version 4.2. The curves are scaled such that the superburst
peak fluxes match at $t\simeq800\,\mathrm{s}$. 

We obtain a measure of the persistent luminosity during the month
preceding and following the superburst from flux measurements obtained
with the PCA on RXTE and the Wide-Field Cameras (WFCs) on-board BeppoSAX,
that are collected in the Multi-Instrument Burst Archive (MINBAR;
e.g., \citealt{Keek2010}). A bolometric correction is available for
several orbits. We use the mean value: $1.4$. By comparing to the
Eddington luminosity for a $1.4\, M_{\odot}$ neutron star with a
$11.2$~km radius and an atmosphere of solar composition (Sect.~\ref{sub:Initial-model-setup}),
we find that the accretion rate was $0.12\,\dot{M}_{\mathrm{Edd}}$
with a root mean-squared deviation of $10\%$.

For models with an accretion rate of $0.10\,\dot{M}_{\mathrm{Edd}}$,
the scaling factor for our simulated curves is approximately $0.42$.
This factor can be explained by a $54\%$ larger distance to the source
than the $6\,\mathrm{kpc}$ that we assumed (\citealt{Galloway2008catalog}).
These models predict a much longer decay time than observed (Fig.~\ref{fig:profile_1636}).
The best fit is provided by a model at higher accretion rate: $Q_{\mathrm{b}}=0.21\,\mathrm{MeV\, nucleon^{-1}}$
and $\dot{M}=0.40\,\dot{M}_{\mathrm{Edd}}$. For those models the
scaling factor is approximately $0.63$, which can be explained by
a $26\%$ larger distance to the source. The best fit model lacks
a precursor burst.

\section{Discussion}

\subsection{Superburst models}

We create $86$ models of a neutron star envelope with accretion of
carbon-rich material in the range of observed accretion rates, assuming
a range of values for the crustal heating parameter $Q_{\mathrm{b}}$
(Fig.~\ref{fig:Mass-accretion-rate}). We follow the thermonuclear
burning of the accreted carbon, which proceeds as flashes (superbursts)
in some cases, whereas at higher values of $Q_{\mathrm{b}}$ burning
becomes stable. We compare the amount of crustal heating of our bursting
models to models of $Q_{\mathrm{b}}$ as a function of $\dot{M}$
(\citealt{Cumming2006}). Only the lowest curve that spans the entire
range of mass accretion rates lies within the range of $Q_{\mathrm{b}}$
where we find unstable carbon burning. This implies that a high neutrino
emissivity of the neutron star core is favored.

Note that superbursting sources mostly accrete hydrogen or helium-rich
material, which create carbon-rich ashes from thermonuclear burning.
By directly accreting the latter composition, we skip the computationally
expensive hydrogen/helium burning, making it possible to simulate
the long superburst recurrence times. Hydrogen/helium burning may
increase the temperature of the envelope, which can be modelled by
an extra contribution to $Q_{\mathrm{b}}$.

Carbon ignition occurs in our models at a column depth $y_{\mathrm{ign}}$
between approximately $10^{10}\,\mathrm{g\, cm^{-2}}$ and $10^{14}\,\mathrm{g\, cm^{-2}}$
(Fig.~\ref{fig:yign_qb}; Fig.~\ref{fig:yign_tign}). The stable
burning models extend to lower $y_{\mathrm{ign}}$ than the bursting
models, because at a given accretion rate stable burning requires
a higher crustal heat flux than unstable burning, which leads to shallower
ignition.

In the relation between the ignition depth (or density) and temperature
(Fig.~\ref{fig:yign_tign}), there is a down turn in the trend for
$y_{\mathrm{ign}}\gtrsim10^{13}\,\mathrm{g\, cm^{-2}}$, which is
due to increased screening of the Coulomb barrier of the carbon ions
at temperatures below $T\lesssim3\cdot10^{8}\,\mathrm{K}$ and densities
in excess of $\rho\gtrsim10^{9}\,\mathrm{g\, cm^{-3}}$ (\citealt{Salpeter1969,Yakovlev2006};
see also, e.g., \citealt{Brown1998}). This is the transition from
thermonuclear burning to pycnonuclear burning, which sets in at temperatures
$T\lesssim10^{8}\,\mathrm{K}$. The ignition conditions are more uncertain
in this regime, because the possible formation of a crystal lattice
may require higher densities for carbon fusion (\citealt{Yakovlev2006}).

When the ignition column depth exceeds $y_{\mathrm{ign}}\gtrsim4\cdot10^{11}\,\mathrm{g\, cm^{-2}}$
we find superbursts to be powerful enough to drive a shock to the
surface (see also \citealt{Weinberg2006sb,Weinberg2007}). At the
start of the thermonuclear runaway, the burning time scale at the
ignition depth becomes shorter than the dynamical time scale. A combustion
wave moves outward, depleting the inner zones of carbon, while initiating
a shock. This detonation phase lasts only a few microseconds, until
the velocity of the combustion wave is sufficiently reduced, and burning
continues to spread as a deflagration. The shock then no longer follows
the burning front, but speeds ahead toward the surface.

\subsection{Decay profile}

\begin{figure}
\includegraphics{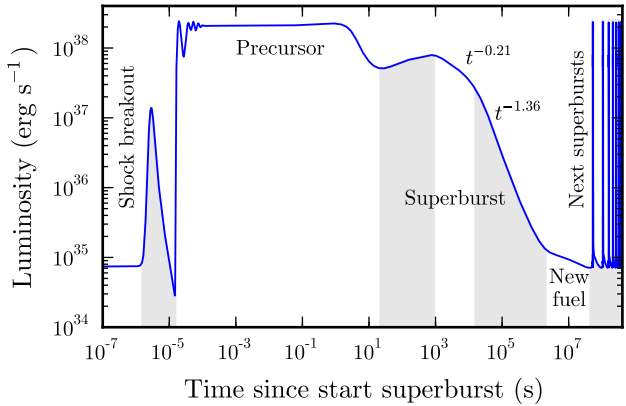}

\caption{\label{fig:schematic}Light curve of the model with $\dot{M}=0.30\,\dot{M}_{\mathrm{Edd}}$
and $Q_{\mathrm{b}}=0.13\,\mathrm{MeV\, nucleon^{-1}}$ , where the
outer zone is at a column depth of $\sim10^{8}\,\mathrm{g\, cm^{-2}}$.
We indicate the different phases of the superburst. Several followup
superbursts are shown.}
\end{figure}
 The superburst light curves consist of a shock breakout, a precursor,
a rise to the `superburst peak', and two power-law decay phases with
$L\propto t^{-0.21}$ and $L\propto t^{-1.36}$ (Fig.~\ref{fig:schematic}).
These power law indices are consistent with the results obtained by
\citet{Cumming2006} and by \citet{Weinberg2007}. \citet{2004CummingMacBeth}
explain that the first power law component is due to radiative cooling
when a cooling wave propagates from the surface inwards (see also
\citealt{Cumming2006}). Once this wave reaches the bottom of the
carbon burning layer, the cooling transitions to the steeper power
law. Hotter models with high accretion rates burn shallower columns,
such that the $L\propto t^{-0.21}$ phase lasts shorter in these models.
For models with the highest values for $\dot{M}$ and $Q_{\mathrm{b}}$,
this phase is absent in the light curve. After the superburst peak
an immediate transition is made to $L\propto t^{-1.36}$.

The $L\propto t^{-1.36}$ ends when burning of newly accreted carbon
starts dominating the light curve, up to the thermonuclear runaway
of the next superburst (Fig.~\ref{fig:schematic}).

Observationally, the duration of the decay is often measured by fitting
an exponential to the light curve. Even though theoretically we expect
the decay to follow a power law, the quality of the data is often
such that an exponential still provides a good fit. Decay times have
been observed between $0.7$ and $6.0$~hours (e.g., \citealt{Keek2008int..work}).
We determine the exponential decay time $t_{\mathrm{exp}}$ for our
light curves as the time it takes for the luminosity to drop from
the peak by one e-fold (Fig.~\ref{fig:tefold_qb}). If this time
falls within the $L\propto t^{-0.21}$ phase, where all models share
the same curve, $t_{\mathrm{exp}}$ is mainly dependent on the peak
luminosity: a higher peak means $\mathrm{e}^{-1}L_{\mathrm{peak}}$
is reached earlier on the $L\propto t^{-0.21}$ curve. This yields
smaller values of $t_{\mathrm{exp}}$ for colder models, because they
have larger peak luminosities. If $t_{\mathrm{exp}}$ reaches into
the $L\propto t^{-1.36}$ phase, $t_{\mathrm{exp}}$ depends less
on the peak, and more on the actual decay of the light curve. Colder
models with lower accretion rates have longer $t_{\mathrm{exp}}$
in this case. The difference in behavior during the two power law
decay phases makes it difficult to use the published observed $t_{\mathrm{exp}}$
to constrain model parameters. It is, therefore, more instructive
to fit a two-component power law decay, if the data quality allows
for it (\citealt{Cumming2006,Keek2008,Kuulkers2010}; Fig.~\ref{fig:profile_1636}).

\subsection{Precursor bursts}

At the start of the thermonuclear runaway a shock travels from the
ignition depth to the surface. The shock breakout is visible in the
light curve as a short bright peak in the light curve, when during
a fraction of a microsecond super-Eddington luminosities are reached
(e.g., Fig.~\ref{fig:atmospheres}). The atmosphere is pushed upwards,
before it falls back on a dynamical time scale of the order of $10^{-5}\,\mathrm{s}$,
and undergoes damped oscillations. The shock and the fall-back deposit
heat for the different models at column depths between $y\simeq10^{8}\,\mathrm{g\, cm^{-2}}$
and $y\simeq10^{10}\,\mathrm{g\, cm^{-2}}$. The cooling of these
layers on thermal time scales of up to approximately $10^{2}\,\mathrm{s}$
is visible as a precursor burst. We find that the peak luminosity
of the shock breakout and the amplitude of the oscillations depends
strongly on the resolution of the atmosphere in the models. The shock
over-pressure is greater closer to the surface (\citealt{Weinberg2007}),
where the density is lower, causing a stronger shock breakout. Note
that not all models produce a precursor: superbursts with $y_{\mathrm{ign}}\lesssim4\cdot10^{11}\,\mathrm{g\, cm^{-2}}$
are not powerful enough to drive a shock. These models still reach
a temperature of $T\simeq5\cdot10^{8}\,\mathrm{K}$ at a column depth
of $y=10^{8}\,\mathrm{g\, cm^{-2}}$. This is too cold for carbon
burning, but it is hot enough to trigger the thermonuclear runaway
of helium burning (e.g., \citealt{Bildsten1998}), leading to a short
hydrogen/helium precursor burst.

Our models confirm the scenario of precursors due to shock heating
(\citealt{Weinberg2007}). \citet{Weinberg2007} created models of
superbursts in a neutron star with a helium-rich envelope. They found
that without the helium, carbon burning at lower depths produces a
weak precursor. In contrast, we find that the shock heating is much
stronger than any carbon burning at this depth, resulting in a bright
precursor from heating alone. Their light curves exhibit a precursor
that starts several seconds after the shock breakout. The shock leaves
a the outer envelope isothermal, delaying the precursor emission by
a thermal time scale. \citet{Weinberg2007} describe that at this
point in the simulation they \emph{assume} a new hydrostatic equilibrium
in the outer layers. Instead, we perform a fully self-consistent calculation
that includes the fall-back of the outer layers, which disrupts the
flat temperature profile, and causes the light curve to change much
faster on a dynamical time scale. Furthermore, most of the energy
of the shock went into the expansion of the outer layers, and is only
converted into heat during the fall-back. This generates, therefore,
not only a precursor more quickly on a dynamical time scale, but also
a much more powerful precursor burst that can last up to $10^{2}\,\mathrm{s}$.

The light travel time around a neutron star with a $10\,\mathrm{km}$
radius is $1.1\cdot10^{-4}\,\mathrm{s}$. Therefore, assuming instantaneous
ignition throughout the entire envelope, any detail in the light curve
at shorter times, such as the shock breakout and the subsequent oscillations,
will be smeared out.

Only for seven superbursts the onset has been observed. Three times
a precursor burst was seen directly at the start of the superburst
(\citealt{Strohmayer2002,Strohmayer2002a,Zand2003}). For two superburst
candidates from GX~17+2 the onset was seen, but because of the quality
of the data only hints of precursors were observed (\citealt{Zand2004}).
The tentative observation of the superburst rise from 4U~1608-522
with HETE-2 did not exhibit a precursor. The data quality allows for
only a weak precursor.

The precursors of 4U~1636-53 and 4U~1254- 69 have a higher peak
flux than the superburst. For 4U~1820-30 the superburst itself was
exceptionally bright, and the precursor's peak flux was approximately
$15\%$ lower. The light curves of the precursor bursts exhibit a
double peak, similar to photospheric radius expansion (PRE) bursts.
For none of these bursts, however, is spectroscopic information available
at sufficient time resolution or of sufficient quality to confirm
the PRE nature of these bursts. Nonetheless, the precursor bursts
in our models all reach the Eddington limit, resulting in PRE.

There is a tentative observation of the onset of the superburst from
4U~1608-522 with the WXM and FREGATE instruments on-board the HETE-2
satellite (\citealt{Keek2008}). At the end of an orbit there is an
increase in the count rate visible for $50\,\mathrm{s}$. No short
precursor can be discerned. The superburst occurred during a transient
outburst, when the accretion rate exceeded $0.10\,\dot{M}_{\mathrm{Edd}}$.
The time-averaged rate in the years before the superburst, however,
was only $0.01\,\dot{M}_{\mathrm{Edd}}$. Our models at low accretion
rates include precursors with durations in excess of $50\,\mathrm{s}$.
This suggests the possibility that the entire flare observed with
HETE-2 was part of the precursor.

\citet{Kuulkers2002ks1731} refer to a burst from KS~1731-260, that
occurred $200\,$s before the superburst rise, as precursor. The $200\,\mathrm{s}$
is much shorter than the typical burst recurrence time for this source
of several hours, as well as shorter than the superburst duration.
The three other precursors, however, occurred immediately prior to
the rise of the superburst. Also, this burst was relatively weak compared
to most other bursts from this source, but several bursts with a similar
peak flux have been observed. Therefore, we suggest that the burst
preceding the superburst of KS~1731-260 is an ordinary burst that
by chance was close to the superburst. Perhaps heating from the stable
carbon burning before the superburst thermonuclear runaway caused
the normal burst to ignite earlier, resulting in a relatively weak
burst.

\subsection{Burning ashes and crustal composition}

Carbon burning and subsequent $\alpha$-capture reactions create $\mathrm{^{20}Ne}$,
$\mathrm{^{24}Mg}$, $\mathrm{^{28}Si}$, and $\mathrm{^{32}S}$ (e.g.,
\citealt{Schatz2003ApJ}). For stable burning models these isotopes
combined with $^{56}$Fe from the ashes of hydrogen/helium bursts
(our accretion composition; e.g., \citealt{Woosley2004}) form the
composition of the outer crust. In superbursting models, photodisintegration
provides more $\alpha$-particles for captures, creating iron group
elements (Fig.~\ref{fig:masses}). Therefore, superbursting neutron
stars have an outer crust composed of mostly iron.

In the bursting models there is a region at a depth lower than the
ignition column depth, where the temperature is insufficient for photodisintegration.
Therefore, carbon burning in this layer creates $\mathrm{^{20}Ne}$,
$\mathrm{^{24}Mg}$, $\mathrm{^{28}Si}$, and $\mathrm{^{32}S}$ (Fig.~\ref{fig:massfractions}).
Note that the most abundant element is iron from the accretion composition.
The next superburst ignites on top of this layer, and these isotopes
burn to iron. A layer enriched in $\mathrm{^{20}Ne}$, $\mathrm{^{24}Mg}$,
$\mathrm{^{28}Si}$, and $\mathrm{^{32}S}$ in principle has a somewhat
lower opacity than a pure iron layer, resulting in a larger ignition
column depth $y_{\mathrm{ign}}$. The layer composition is, however,
dominated by iron, which limits the changes in opacity. We do not
find a difference between $y_{\mathrm{ign}}$ of the first superburst
in a series and subsequent bursts that can be attributed to such a
compositional inertia effect (cf. \citealt{Woosley2004}).

\subsection{Recurrence times\label{sub:Recurrence-times}}

The bursting models exhibit recurrence times of several days up to
thousands of years (Fig.~\ref{fig:trecur_qb}). We compare these
to the three cases where more than one superburst has been observed
from the same source. For these sources we derive the time averaged
mass accretion rate from the persistent X-ray luminosity as reported
in the MINBAR catalog (e.g., \citealt{Keek2010}), expressed in units
of the Eddington limited rate for our choice of neutron star parameters
(Sect.~\ref{sub:Initial-model-setup}). The typical uncertainty in
measurements of the accretion rate are several tens of percents. Due
to the presence of frequent data gaps, the observed recurrence times
have to be regarded as upper limits (e.g., \citealt{Keek2010} for
the case of bursts with short recurrence times).

Ser~X-1 has a mean mass accretion rate of $0.31\,\dot{M}_{\mathrm{Edd}}$,
and superbursts observed at least 2.4~years apart (\citealt{Cornelisse2000,Kuulkers2009ATel}).
The models with $0.30\,\dot{M}_{\mathrm{Edd}}$ have a maximum recurrence
time of $t_{\mathrm{recur}}=1.7\,\mathrm{years}$ ($2.1$~years for
a gravitational redshift of $z+1=1.26$), whereas models with $0.20\,\dot{M}_{\mathrm{Edd}}$
reproduce the observed $t_{\mathrm{recur}}$. Therefore, the accretion
rate is somewhat lower than observationally inferred, or the actual
$t_{\mathrm{recur}}$ of this source is shorter than the upper limit.

Four superbursts have been observed from GX~17+2; two only 15~days
apart (\citealt{Zand2004}). The mass accretion rate is unusually
high for a bursting source: on average $1.21\,\dot{M}_{\mathrm{Edd}}$.
The models with a mass accretion rate of $1.00\,\dot{M}_{\mathrm{Edd}}$
exhibit recurrence times up to 11~days (13~days for $z+1=1.26$).
It is likely that models with a smaller accretion rate of $\sim0.9\,\dot{M}_{\mathrm{Edd}}$
will produce the observed recurrence time. This rate is within the
uncertainty of the observed accretion rate.

The shortest time interval between two observed superbursts of 4U~1636-53
is 1.1~year (\citealt{Wijnands2001sb,Strohmayer2002a,2004Kuulkers,Kuulkers2009ATel}).
The time averaged mass accretion rate of 4U~1636-53 is $0.12\,\dot{M}_{\mathrm{Edd}}$,
but the models close to this rate produce a recurrence time of at
least $7\,\mathrm{years}$ ($9$~years for $z+1=1.26$). Part of
this problem may be explained by the uncertainties in the measured
mass accretion rate of several tens of percents. Other explanations
for this discrepancy may be found in our assumptions for the carbon
mass fraction and of the effective gravity in the neutron star envelope.

For nine sources, most of which with accretion rates close to $0.10\,\dot{M}_{\mathrm{Edd}}$,
a lower limit to the superburst recurrence time was derived from the
BeppoSAX Wide-Field Camera (WFC) data (\citealt{kee06}). The average
lower limit of $60$~days is indicated in Fig.~\ref{fig:trecur_qb},
and is consistent with the model results for the sources with mass
accretion rates up to $0.5\,\dot{M}_{\mathrm{Edd}}$.

The same recurrence time can be reproduced by models with a mass accretion
rate that varies by several tens of percents, and different values
of $Q_{\mathrm{b}}$. This spread in mass accretion rate is of the
order of the uncertainty in the observed rate, which makes it difficult
to constrain $Q_{\mathrm{b}}$ from the observed recurrence times.

\subsection{Comparison to 4U~1636-53}

The most detailed superburst light curves have been observed with
the PCA on RXTE: one from 4U~1636-53 (\citealt{Strohmayer2002a})
and one from 4U~1820-30 (\citealt{Strohmayer2002}). The latter source
is an ultra-compact binary system, which implies that the accreted
material is hydrogen-deficient. The material that burns in the superburst
is thought to be more carbon-rich than the composition that we assumed
in our calculations. Its superburst is atypical, as the superburst
peak reached the Eddington limit and exhibited radius expansion. For
these reasons we do not compare to this source, and focus our attention
on the superburst from 4U~1636-53.

The superburst from this source started with a short precursor that
showed behavior consistent with photospheric radius expansion, and
that decayed on a 10~s time scale. The superburst decay was observable
for $5.5\,\mathrm{hours}$. The average mass accretion rate of this
source over the past decade is $0.12\,\dot{M}_{\mathrm{Edd}}$ (Sect.~\ref{sub:Recurrence-times}).
The models close to this rate exhibit a much longer decay time scale,
even for high crustal heating (Fig.~\ref{fig:profile_1636} top).
Also, the precursors of these models last longer than observed. The
decay is best fit by a model with a four times higher accretion rate
of $\dot{M}=0.40\,\dot{M}_{\mathrm{Edd}}$ and with $Q_{\mathrm{b}}=0.21\,\mathrm{MeV\, nucleon^{-1}}$
(Fig.~\ref{fig:profile_1636} middle). This models, however, does
not have a precursor, and the part leading up to the superburst peak
is not well reproduced. This may be due to the fact that the atmosphere
of 4U~1636-53's neutron star is probably hydrogen-rich, instead of
carbon-rich as we assumed. While our modeled burst is not powerful
enough to heat the atmosphere by a shock, the temperature in the atmosphere
is high enough, $T\simeq6\cdot10^{8}\,\mathrm{K}$, to ignite a hydrogen/helium
precursor burst.

If we were to assume that $\dot{M}=0.40\,\dot{M}_{\mathrm{Edd}}$
is a good approximation, Fig.~\ref{fig:profile_1636} (bottom) illustrates
how sensitive the decay depends on the crustal heating parameter $Q_{\mathrm{b}}$.

The discrepancy between the observed light curve and the models at
$0.10\,\dot{M}_{\mathrm{Edd}}$ adds to the problem we noted earlier
that the recurrence times predicted by the models with $0.10\,\dot{M}_{\mathrm{Edd}}$
are substantially longer than the observational upper limit (Fig.~\ref{fig:trecur_qb}).
The answer to this problem may lie in the fact that we only considered
one carbon fraction for the envelope, and one value for the effective
gravity. Variations of these parameters could yield the shallower
ignition implied by our models. Another possibility are multi-dimensional
effects that we cannot model in our one-dimensional code. Pulsations
at the spin frequency of the neutron star were observed during $800\,\mathrm{s}$
close to the superburst peak (\citealt{Strohmayer2002a}). This indicates
that the emission was anisotropic during a quite long period after
superburst ignition, which hints at the presence of multi-dimensional
effects.

\section{Conclusions}

To study carbon flashes (superbursts), we constructed 86 one-dimensional
multi-zone models of the envelope of a neutron star that accretes
carbon-rich material, for different mass accretion rates and amounts
of crustal heating. These are the first such models that were constructed
by following the accumulation of the fuel layer and the thermonuclear
burning of carbon during a series of superbursts. The stability of
carbon burning is investigated as a function of the amount of crustal
heating. We reproduced the two-component power-law decay (\citealt{2004CummingMacBeth}).
Not all models, however, exhibit the first component: the hotter models
at higher mass accretion rates show a direct transition from the superburst
peak to the second (steeper) power-law component.

The superburst ashes that form the outer crust are primarily composed
of iron. Carbon burning higher up in the envelope produces isotopes
with mass numbers around $30$. The next superburst ignites in this
layer. We do not find a compositional inertia effect, as seen for
hydrogen/helium bursts (\citealt{Woosley2004}), because the layer
composition is dominated by the accreted fraction of iron. In case
of a larger carbon fraction of the fuel layer, however, such an effect
may become important.

We obtain a precursor burst due to shock heating, similar to \citet{Weinberg2007}.
We find that heating by the shock and the fall-back of expanded layers
is sufficient for a strong precursor, that starts approximately $10^{-5}\,\mathrm{s}$
after the shock breakout, instead of seconds. For hot models, at large
accretion rates, the superburst is not powerfull enough to generate
a shock and thus a precursor from shock heating. At low accretion
rates the precursors have durations as long as $10^{2}\,\mathrm{s}$.
This may explain the lack of a short precursor in the observation
of the onset of the superburst from 4U~1608-522.

Comparing the model light curves to the superburst observations with
the PCA on RXTE of 4U~1636-53, the models at the observationally
inferred mass accretion rate overpredict the superburst duration and
the recurrence time. The best agreement is found with models at a
three times higher accretion rate. The discrepancy may be caused by
the values we assumed for the carbon fraction in the ocean and the
effective gravity. This can be studied further with one-dimensional
models. Alternatively, it can be a sign of multi-dimensional effects,
where a higher local accretion rate is responsible the observed behavior.

We studied the dependence of observables, such as the recurrence time
and the shape of the light curve, on the amount of crustal heating
$Q_{\mathrm{b}}$. While we show that these observables can depend
strongly on $Q_{\mathrm{b}}$, the example of 4U~1636-53 indicates
that without a good agreement of the behavior as a function of mass
accretion rate, it is difficult to constrain $Q_{\mathrm{b}}$.

\acknowledgements{The authors thank Andrew Cumming for providing crustal heating models,
Ke-Jung Chen for helpful discussion, and the University of Minnesota
Supercomputing Institute for support. LK is supported by the Joint
Institute for Nuclear Astrophysics (JINA; grant PHY02-16783), a National
Science Foundation Physics Frontier Center. AH acknowledges support
from the DOE Program for Scientific Discovery through Advanced Computing
(SciDAC; DE-FC02-09ER41618) and by the US Department of Energy under
grant DE-FG02-87ER40328.}

\bibliographystyle{apj}
\bibliography{apj-jour,sbkepler}

\appendix

\section{Model light curves}

\begin{figure*}
\includegraphics{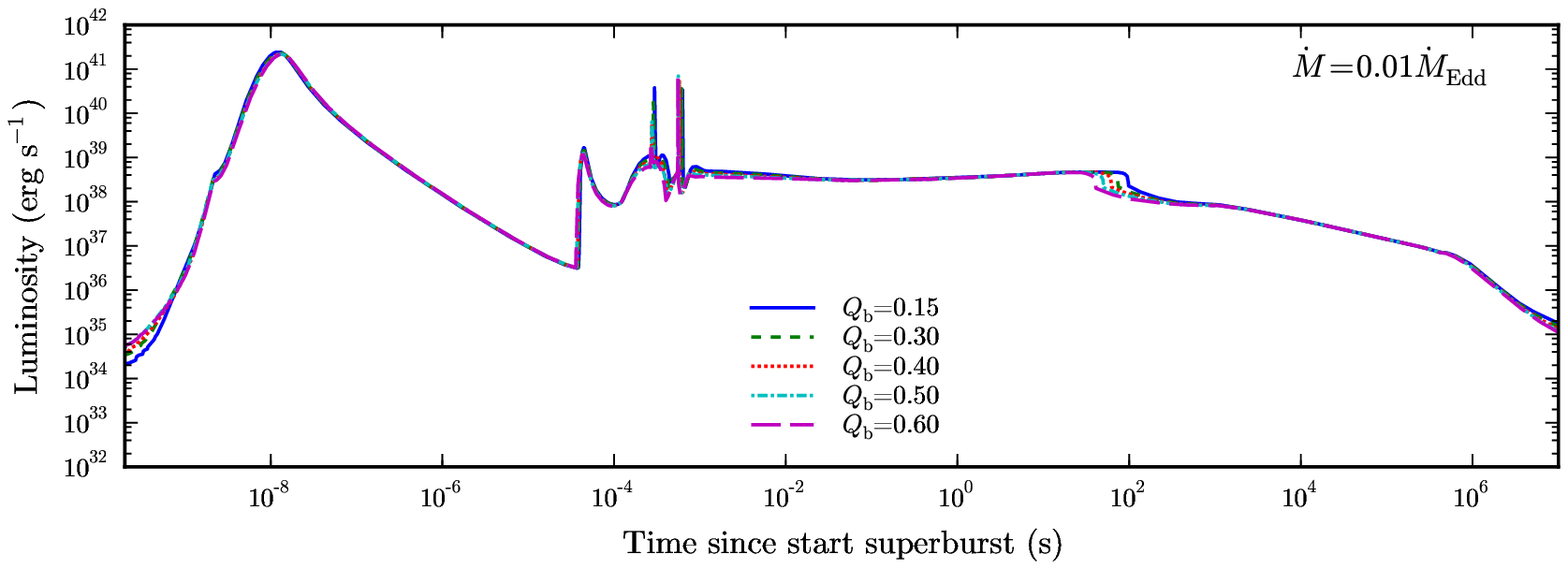}

\includegraphics{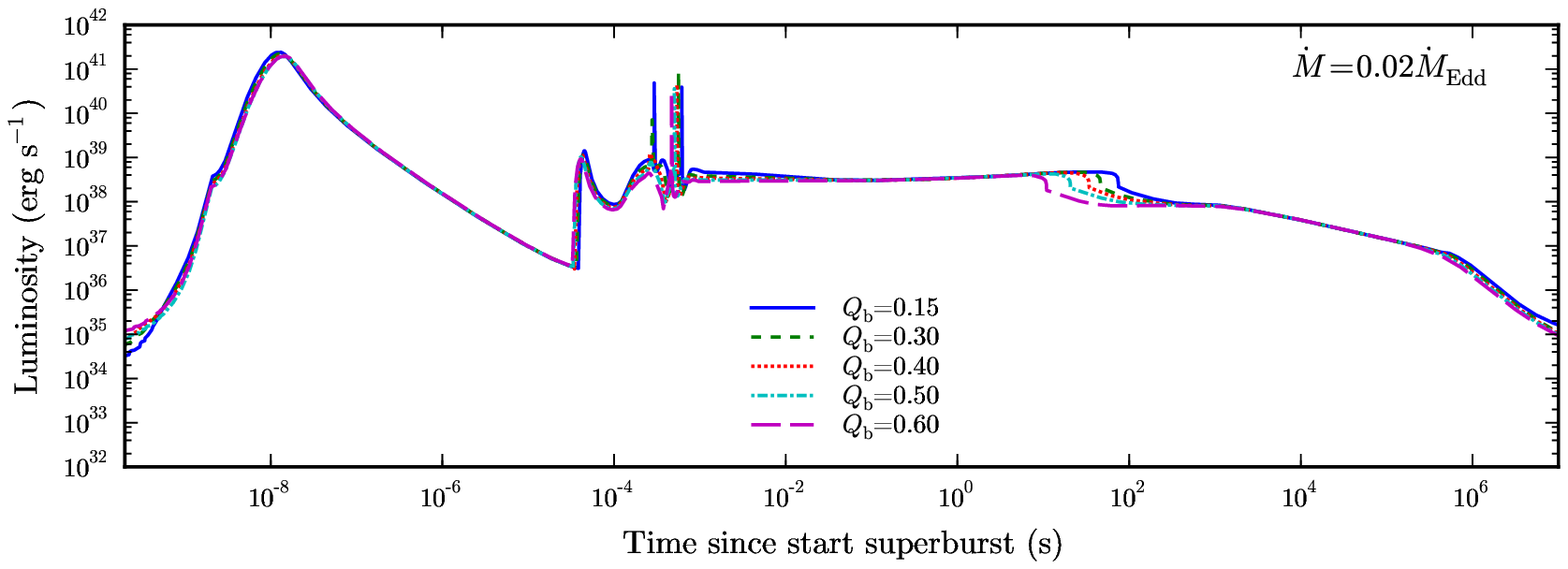}

\includegraphics{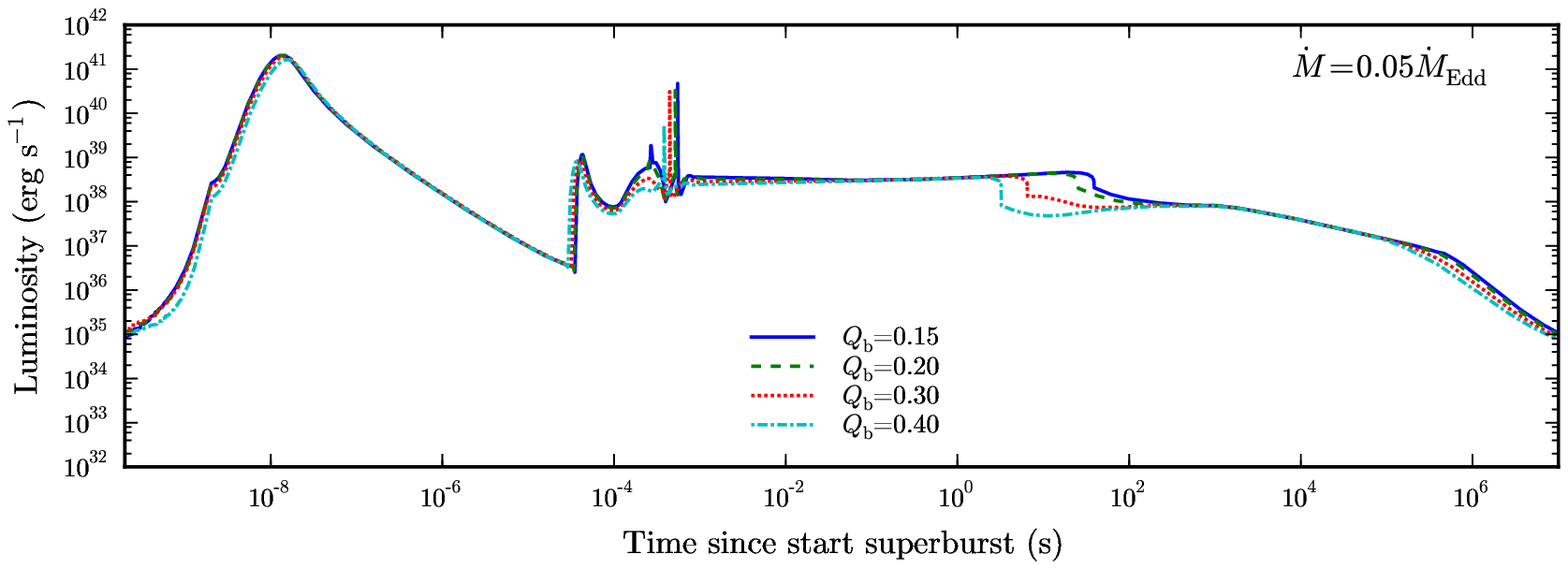}

\includegraphics{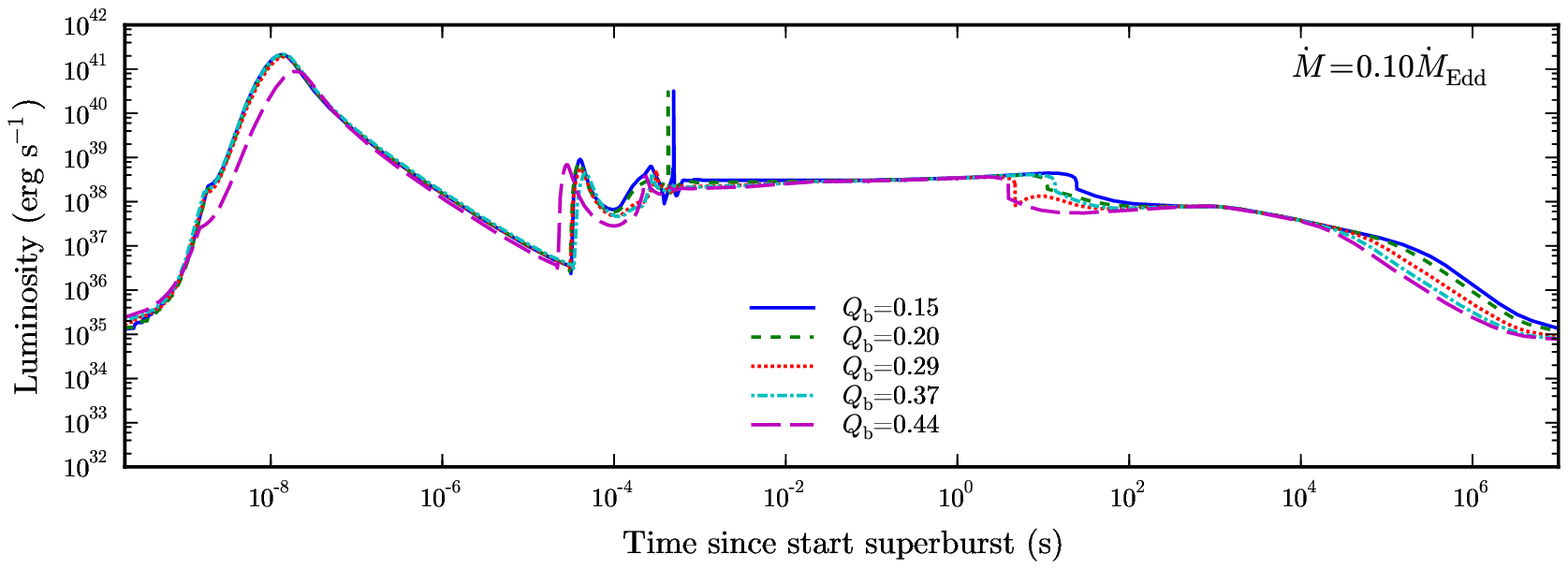}

\caption{\label{fig:all-light-curves}Model light curves. Each panel shows
the light curves resulting from models with a certain mass accretion
rate $\dot{M}$, at different values of the crustal heating parameter
$Q_{\mathrm{b}}$. Continued on next page.}
\end{figure*}

We present all light curves resulting from the models indicated in
Fig.~\ref{fig:Mass-accretion-rate} (Fig.~\ref{fig:all-light-curves}).
The curves do not contain corrections for the gravitational redshift
near neutron stars. See Section~\ref{sub:Light-curve} for further
details.

\begin{figure}

\includegraphics{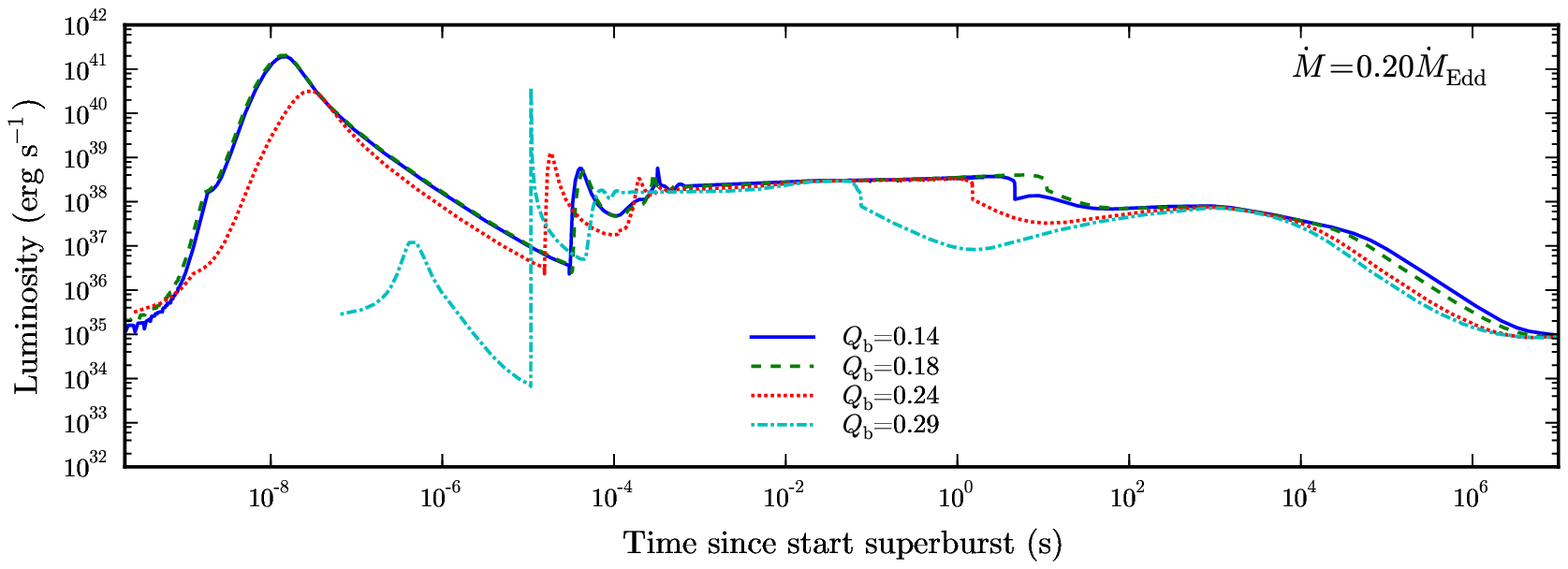}

\includegraphics{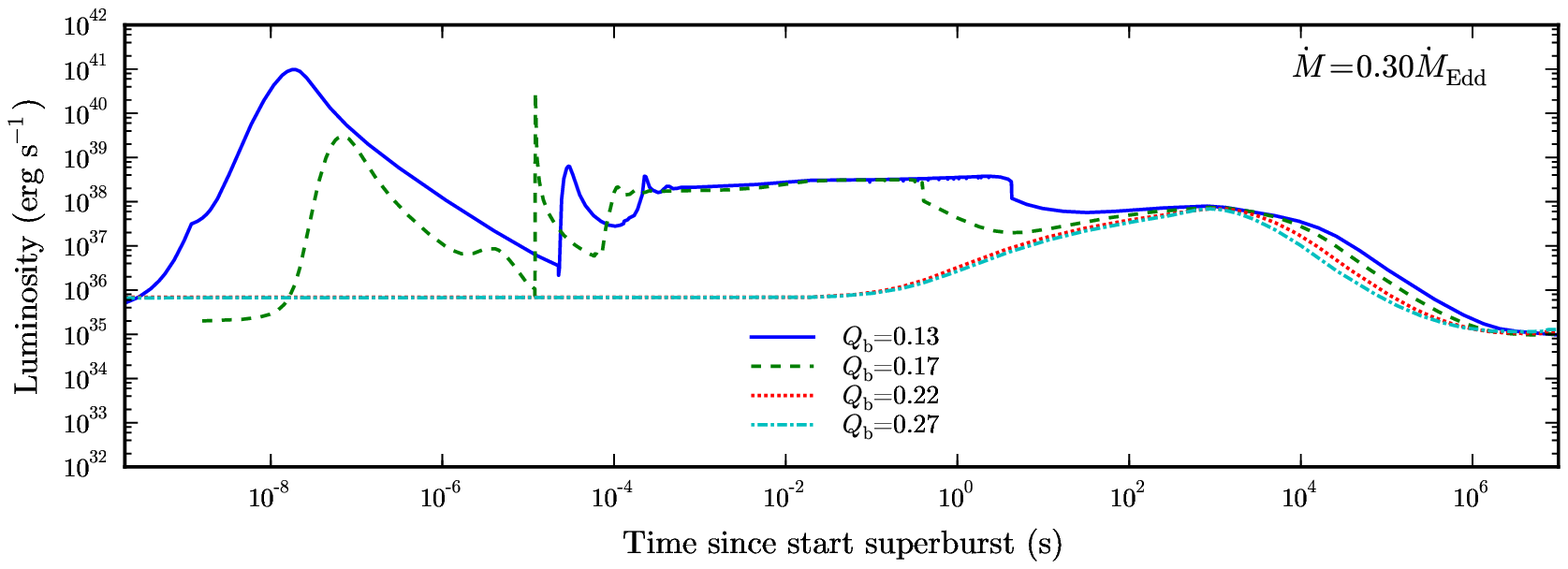}

\includegraphics{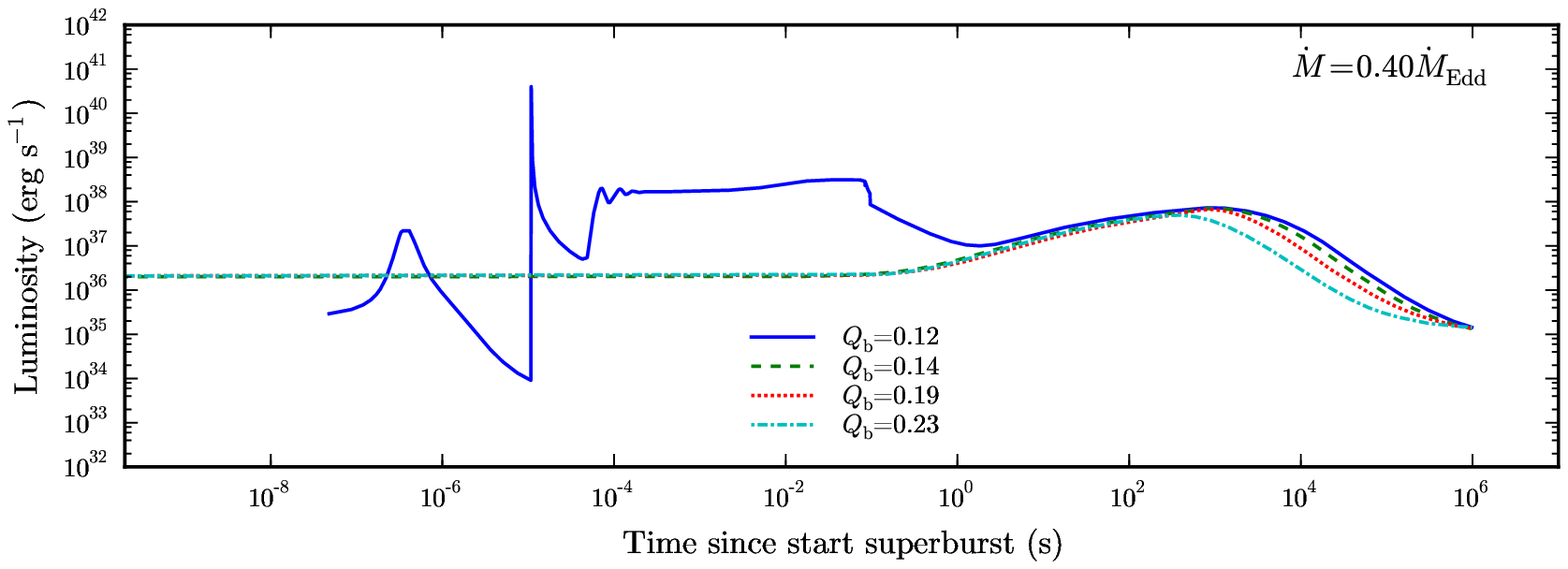}

\includegraphics{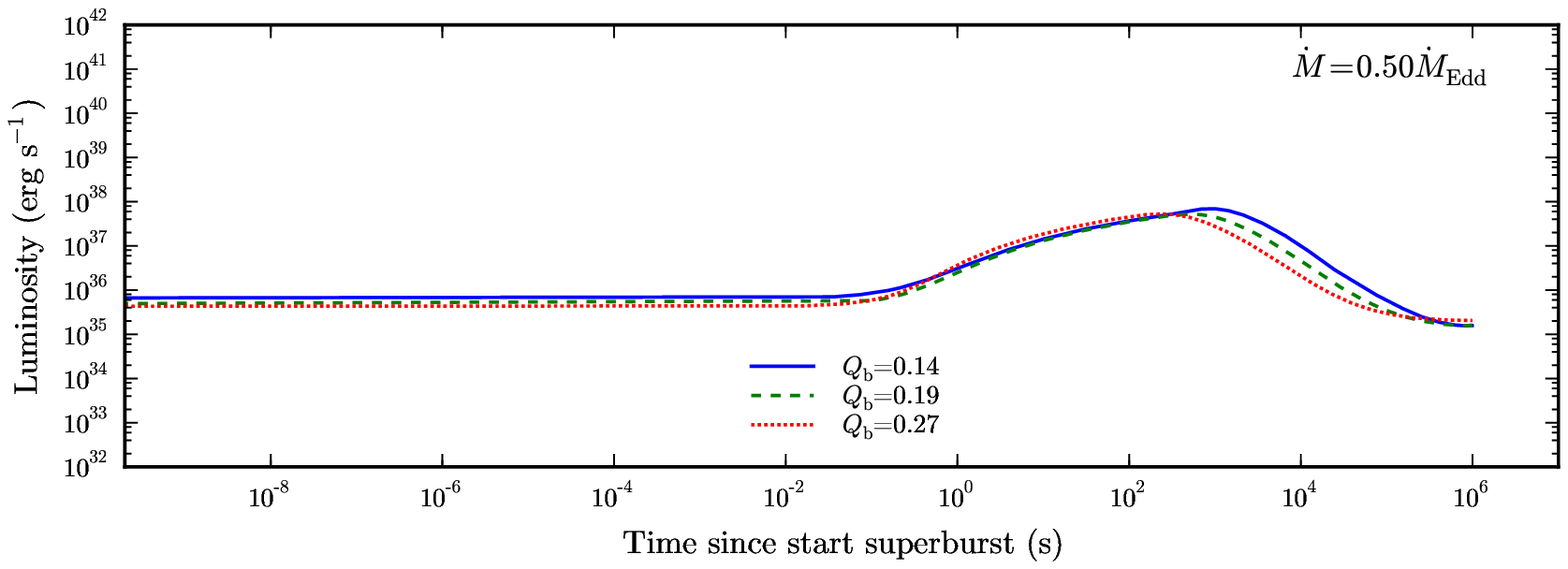}

\textsc{Fig.~\ref{fig:all-light-curves} continued} Continued on
next page.
\end{figure}
\begin{figure}

\includegraphics{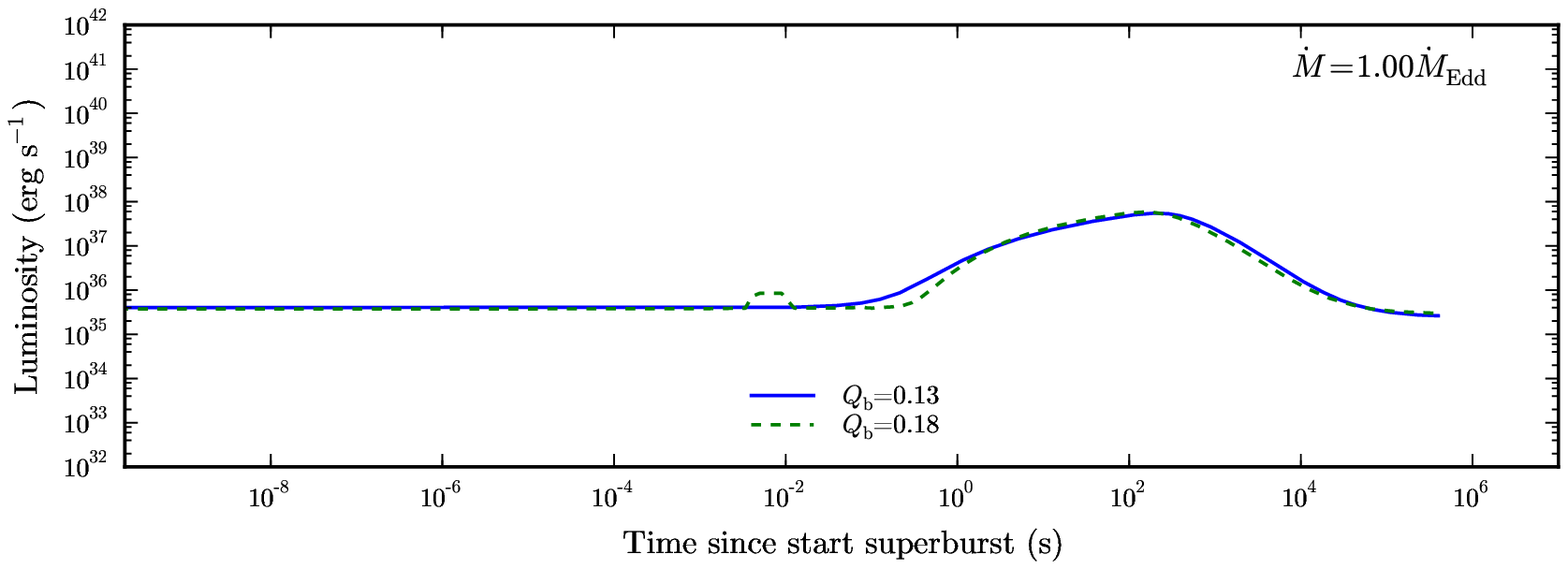}

\textsc{Fig.~\ref{fig:all-light-curves} continued}
\end{figure}

\section{General relativistic corrections}
\label{sec:General-relativistic-corrections}

\newcommand{\gr}{{g_\mathrm{rel}}} \newcommand{\gri}{{g_\mathrm{rel,1}}} \newcommand{\ri}{{r_\mathrm{1}}} \newcommand{\zi}{{z_\mathrm{1}}} \newcommand{\Mi}{{M_\mathrm{1}}} \newcommand{\Mf}{{M_\infty}} \newcommand{\Vi}{{{\cal V}_\mathrm{1}}} \newcommand{\Ledd}{{L_\mathrm{Edd}}} \newcommand{\Leddi}{{L_\mathrm{Edd,1}}} \newcommand{\Leddf}{{L_{\mathrm{Edd},\infty}}} \newcommand{\Msun}{{M_{\odot}}} \newcommand{\Mdot}{{\dot{M}}} \newcommand{\Mdoti}{{\dot{M}_1}} \newcommand{\Mdotf}{{\dot{M}_\infty}} \newcommand{\Ri}{{r_1}} \newcommand{\Rf}{{r_\infty}} \newcommand{\Li}{{L_1}} \newcommand{\Lf}{{L_\infty}} \newcommand{\Lacc}{{L_\mathrm{acc}}} \newcommand{\Lacci}{{L_\mathrm{acc,1}}} \newcommand{\Laccf}{{L_{\mathrm{acc},\infty}}}
\newcommand{\Laccif}{{L_{\mathrm{acc},1,\infty}}}

\subsection{Introduction}

The thin surface layer of a neutron star in which the X-ray burst
occurs can locally be treated in Newtonian physics for $\Delta r\ll r$,
where $r$ is the radius of the star and $\Delta r$ is the thickness
of the surface layer. For neutron stars general relativistic (GR)
effects are important, but the Kepler code, which is employed in this
study, uses Newtonian gravity. Therefore, we need to correct for GR
effects to allow for proper interpretation and comparison with observations.
This comprises two types of corrections: first, identifying the neutron
star masses and radii that give same relativistic gravitational acceleration
as the Newtonian acceleration that was used from the input mass and
radius. Second, correct the results for time dilatation (of the light
curve and decrease of the accretion rate; $\sim(1+z)$, where $z$
is the gravitational redshift) and weakening of luminosity ($\sim(1+z)^{2}$).

\subsection{Translating Newtonian to GR}

When only Newtonian gravity is used in the calculation, neglecting
the strengthening of gravity by a factor $(1+z)$, the result can
still be interpreted as that of a star with a different mass, $\Mi$,
and radius, $\ri$, and correspondingly adjusted (smaller) redshift,
$\zi$, such that the GR acceleration equals the Newtion acceleration
in the calculation. Below the scaling laws for interpreting mass and
radius are derived.

\citet{Thorne1977} gives the volume redshift factor 
\begin{equation}
\mathcal{V}=1\left/\sqrt{1-2GM/(c^{2}r)}\right.,\label{z}
\end{equation}
 which will be called $1+z$ here. Relativistic gravitational acceleration
is given by (e.g., \citealt{Woodhouse2007}) 
\begin{equation}
\gr=-(1+z)GM\left/r^{2}\right.=GM\left/\left(r^{2}\sqrt{1-2GM/(c^{2}r)}\right)\right..\label{grel}
\end{equation}

We now define a radius $\ri$ and an actual gravitational mass, $\Mi$,
such that the GR gravity at this point equals the Newtonian gravity,
$g$, at radius $r$, i.e., $\gri\stackrel{!}{=}g$: 
\begin{eqnarray}
GM\left/r^{2}\right.\stackrel{!}{=}G\Mi\left/\left(\ri^{2}\sqrt{1-2G\Mi\left/\left(c^{2}\ri\right)\right.}\right)\right.\;.\label{EQ:eq}
\end{eqnarray}
 This can be rewritten as 
\begin{eqnarray}
\varphi^{2}+2\varphi\zeta\xi^{3}-\xi^{4}=0\qquad\mbox{with}\qquad\xi=\ri/r\;,\qquad\varphi=\Mi/M\;,\qquad\zeta=GM\left/\left(c^{2}r\right)\right.\;,\label{reduced}
\end{eqnarray}
 where $\zeta$ is the \emph{gravitational radius} of the original
problem.

\subsubsection{Given Mass\label{sub:Given-Mass}}

Assuming a mass $\Mi=\varphi M$, the physical solution of this 4$^{\mathrm{th}}$
order equation for $\xi$ is given by 
\begin{eqnarray}
\xi=\frac{\zeta\,\varphi}{2}\left(1+\sqrt{1-A}+\sqrt{2+A+2/\sqrt{1-A}}\right)\nonumber \\
A=\sqrt[3]{2/9}\left(B^{2}/\varphi^{2}-2\sqrt[3]{6}\right)\left/\left(B\zeta^{2}\right)\right.\;,\qquad B=\sqrt[3]{9\,\zeta^{2}\,\varphi^{4}+{\sqrt{3}}\,\varphi^{3}\,{\sqrt{16+27\,\zeta^{4}\,\varphi^{2}}}}\;.\label{xi}
\end{eqnarray}
 The radius $\ri$ at which the GR gravitational acceleration matches
the Newtonian one for the assumed radius, $r$, is thus given by $\ri=\xi r$.
The redshift factor $\zi$ for radius $\ri$ and mass $\Mi$ is given
by 
\begin{equation}
1+\zi=1\left/\sqrt{1-2G\Mi/(c^{2}\ri)}\right.=1\left/\sqrt{1-2\zeta\varphi/\xi}\right.\;.\label{z1}
\end{equation}
Using Equation~(\ref{EQ:eq}) one obtains 
\begin{equation}
\xi^{2}/\varphi=1+\zi\;.\label{z1a}
\end{equation}

Using these relations , the light curve for an observer at infinity
has to be time dilated by $1+\zi$. Due to the larger radius, the
surface area is increased by $\xi^{2}=\varphi(1+\zi)$ and thus luminosity
has to be scaled by $\xi^{2}/(1+\zi)^{2}$, i.e., decreased by a factor
$(1+\zi)/\varphi$. Similarly, the apparent accretion rate for an
observer at infinity scales as $\xi^{2}/(1+\zi)=\varphi$, that is,
does not need to be modified if $M=\Mi$. For a NS with $1.4\,\Msun$
(gravitational) mass and 10\,km Newtonian model radius and assuming
$M=\Mi$, i.e., $\varphi=1$, $\zeta=0.206666$, one obtains $\xi=1.12176$
and $\zi=0.25835$.

\subsubsection{Given Radius}

On the other hand, if we know the true radius, $\ri=\xi r$, we can
determine the mass corresponding to the gravity we used by solving
Eq.\ (\ref{reduced}) for $\varphi$ and then use Eq.~(\ref{z1})
or (\ref{z1a}) to determine $\zi+1$, 
\begin{equation}
\varphi=\Mi/M=\zeta\xi^{3}\left(\sqrt{1+1\left/\left(\xi^{2}\zeta^{2}\right)\right.}-1\right)\;.\label{phi}
\end{equation}
 For our parameters we then obtain $\xi=1$, $\varphi=0.81440$, i.e.,
$\Mi=1.1401630\,\Msun$, $\zi=0.22789465$.

\subsubsection{Minimal Adjustment}

Alternatively, we could search for a minimum deviation of both $\varphi$
and $\xi$ from $1$, that is, setting $\xi=1/\varphi$ in Eq.~(\ref{reduced}):
\begin{equation}
\varphi^{6}+2\zeta\varphi^{2}-1=0
\end{equation}
 with the physical solution 
\begin{equation}
\varphi^{2}=C/6-4\,\zeta/C\qquad\mbox{where}\qquad C=\sqrt[3]{108+4\,\sqrt{729+864\,\zeta^{3}}}\;.
\end{equation}
 For our parameters we then obtain $\xi=1/\varphi=1.076353$, i.e.,
$\zi=0.246993$, $\Mi=1.30069\,\Msun$, and $\ri=10.764\,$km.

\subsection{Accretion and Eddington Luminosity}

\subsubsection{Eddington Luminosity}

The Eddington luminosity is determined by gravitational acceleration
being balanced by radiation pressure on electrons: $\Ledd=4\pi r^{2}gc/\kappa$,
where $\kappa$ is the opacity. In Newtonian approximation this computes
to 
\begin{equation}
\Ledd=4\pi c\, GM/\kappa.
\end{equation}
This is also the Eddington luminosity `at infinity', as there is no
redshift in the Newtonian case. In the frame of (corrected) stellar
surface, the Eddington luminosity taking into account GR gravity is
given by 
\begin{equation}
\Leddi=\left(1+\zi\right)4\pi c\, G\Mi/\kappa=\varphi\left(1+\zi\right)\Ledd\;.
\end{equation}
This is the same as the scaling factor $\xi^{2}$ for any luminosity
(Sect.~\ref{sub:Given-Mass}).

\subsubsection{Accretion Luminosity}

We summarize the scaling laws for mass, radius, accretion rate and
luminosity: 
\begin{eqnarray}
\Mi=\varphi M=\Mf\;,\\
\Ri=\xi r=\Rf/\left(1+\zi\right)\;,\qquad\Rf=\left(1+\zi\right)\Ri=\xi\left(1+\zi\right)r\;,\\
\Li=\xi^{2}L=\left(1+\zi\right)^{2}\Lf\;,\qquad\Lf=\Li/\left(1+\zi\right)^{2}=\xi^{2}L/\left(1+\zi\right)^{2}\;,\\
\Mdoti=\xi^{2}\Mdot=\left(1+\zi\right)\Mdotf\;,\qquad\Mdotf=\Mdoti/\left(1+\zi\right)=\xi^{2}\Mdot/\left(1+\zi\right)\;.
\end{eqnarray}
 Note that for $\varphi=1$ Eq.~(\ref{z1a}) leads to $\Mdotf=\Mdot$.

The accretion luminosity we define by $\Lacc=-\Mdot\phi$, where $\Mdot$
is the accretion rate and $\phi$ the gravitational potential. In
the Newtonian approximation $\phi=-GM/r^{2}$. For this case we have
an accretion luminosity and the ratio of accretion luminosity to Eddington
luminosity given by 
\begin{eqnarray}
\Lacc=\Mdot GM/r\;,\qquad\Lacc/\Ledd=\Mdot\kappa/4\pi cr\;.
\end{eqnarray}
In GR the gravitational potential is given by $\phi=-c^{2}z/(1+z)$
(from $\gr=\partial_{r}\phi$, Eq.~\ref{z}, \ref{grel}; \citealt{Misner1973grav.book}
\S25.5). Using corrected mass and radius, one obtains 
\begin{eqnarray}
\Lacci=\Mdoti c^{2}\zi/(1+\zi)=\Mdotf c^{2}\zi=\Mdot\varphi c^{2}\zi\;,\\
\frac{\Lacci}{\Leddi}=\frac{\Mdoti\kappa c\zi}{4\pi G\Mi\left(1+\zi\right)^{2}}=\frac{\Mdotf\kappa c\zi}{4\pi G\Mf(1+\zi)}=\frac{\Mdot\xi^{2}\kappa c\zi}{4\pi GM\varphi\left(1+\zi\right)^{2}}=\frac{\Mdot\kappa c\zi}{4\pi GM\left(1+\zi\right)}\;,
\end{eqnarray}
 where we took advantage of Eq.~(\ref{z1a}). For an observer at
infinity, the accretion luminosity is reduced by $\left(1+\zi\right)^{2}$
: 
\begin{eqnarray}
\Laccf=\Lacci/\left(1+\zi\right)^{2}=\Mdoti c^{2}\zi/\left(1+\zi\right)^{3}=\Mdotf c^{2}\zi/\left(1+\zi\right)^{2}=\Mdot\varphi c^{2}\zi/\left(1+\zi\right)^{2}\;,
\end{eqnarray}
Thus we obtain the following scaling relations 
\begin{eqnarray}
\frac{\Lacci}{\Leddi}=\frac{\Laccf}{\Leddf}=\frac{c^{2}r\zi}{GM(1+\zi)}\frac{\Lacc}{\Ledd}=\frac{\zi}{\zeta(1+\zi)}\frac{\Lacc}{\Ledd}\;,
\end{eqnarray}
 That is, for our example and using $\varphi=1$ the ratio of accretion
rate relative to Eddington accretion rate scales by $0.9934$ to the
`GR corrected frame', both at the neutron star surface and in the
frame of the observer.

\subsection{Limiting Neutron Star Properties}

Finally, if the entire light curve can be fit accurately enough to
observations that both gravity and redshift ($\zi$) are well determined,
we can compute the (gravitational) mass and radius of the neutron
star. Using the definitions 
\begin{equation}
\Vi=1+\zi\qquad\mbox{and}\qquad\gamma=1-1/\Vi^{2}=2\zeta\varphi/\xi\;,
\end{equation}
 we obtain 
\begin{eqnarray}
\varphi=\gamma^{2}\left/\left(4\zeta^{2}\sqrt{1-\gamma}\right)\right.=\left(\Vi^{2}-1\right)^{2}\left/\left(4\Vi^{3}\zeta^{2}\right)\right.=\zi^{2}\left(\zi+2\right)^{2}\left/\left(4\zeta^{2}\left(\zi+1\right)^{3}\right)\right.\;,\nonumber \\
\xi=\gamma\left/\left(2\zeta\sqrt{1-\gamma}\right)\right.=\left(\Vi^{2}-1\right)\left/\left(2\zeta\Vi\right)\right.=\zi\left(\zi+2\right)\left/\left(2\zeta\left(\zi+1\right)\right)\right.\;.
\end{eqnarray}
 More generally, mass, $\Mi$ and radius, $\ri$, are obtained from
gravitational acceleration, $g$, and redshift, $\zi$, by 
\begin{equation}
\Mi=c^{4}\zi^{2}(\zi+2)^{2}/(4Gg(\zi+1)^{3})\;,\qquad\ri=c^{2}\zi(\zi+2)/(2g(\zi+1))\;.\label{gz}
\end{equation}

Of course, such a determination would require that all degeneracy
of model gravity with accretion rate and metallicity would be removed
and fitting of the light curve is reliable (and non-degenerate) with
respect to nuclear data, opacities, equation of state, multidimensional
effects, magnetic fields, etc.

\end{document}